\documentclass[a4paper, 12 pt, reqno]{article}
\usepackage[utf8]{inputenc}
\usepackage{enumitem}
\usepackage{amssymb}
\usepackage{amsthm}
\usepackage{mathtools}
\usepackage{caption}
\usepackage{subcaption}
\usepackage{bbold}
\usepackage{xcolor}
\usepackage{dirtytalk}
\usepackage{multirow}
\usepackage{graphicx}
\usepackage{lmodern,textcomp}
\usepackage{placeins}
\usepackage{indentfirst}
\usepackage[left=3cm,right=3cm,top=2.5cm,bottom=4cm]{geometry}
\usepackage{comment}
\usepackage{float}
\usepackage{natbib}
\usepackage[colorlinks=true,allcolors=blue]{hyperref}
\usepackage[nameinlink,noabbrev,capitalize]{cleveref}
\usepackage{authblk}
\usepackage{array}
\makeatletter
\let\@fnsymbol\@arabic
\makeatother

\title{\textcolor{blue}{Flexible and Context-Specific AI Explainability: \\ A Multidisciplinary Approach}\thanks{We would like to thank Thomas Bonald, Jean-Louis Dessalles, Juliette Mattioli, Tanya Perelmuter, Fabian Suchanek, Xavier Vamparys and Tiphaine Viard for insightful discussions during the preparation of the manuscript. This research has been conducted within the Research Chair ”Digital Finance” under the aegis of the Risk Foundation, a joint initiative by Le Groupement Cartes Bancaires CB, la Banque Postale, Cartes Bancaires and Telecom Paris and University of Paris 2 Panthéon-Assas, the Research Chair Explainable Artificial Intelligence for Anti-Money Laundering, and the Research Chair Data Science and Artificial Intelligence for Digitalized Industry and Services.}}

\date{\today}

\author[1]{Valérie Beaudouin}
\author[2]{Isabelle Bloch}
\author[2]{David Bounie}
\author[1]{Stéphan Clémençon}
\author[1]{Florence d’Alché-Buc}
\author[1]{James Eagan}
\author[2]{Winston Maxwell}
\author[1]{Pavlo Mozharovskyi}
\author[1]{Jayneel Parekh}

\affil[1]{LTCI, Telecom Paris, Institut Polytechnique de Paris}
\affil[2]{i3, CNRS, Telecom Paris, Institut Polytechnique de Paris}

\begin{document}

\maketitle
\baselineskip0.7cm

\begin{abstract}
The recent enthusiasm for artificial intelligence (AI) is due principally to advances in deep learning. Deep learning methods are remarkably accurate, but also opaque, which limits their potential use in safety-critical applications. To achieve trust and accountability, designers and operators of machine learning algorithms must be able to explain the inner workings, the results and the causes of failures of algorithms to users, regulators, and citizens. The originality of this paper is to combine technical, legal and economic aspects of explainability to develop a framework for defining the "right" level of explainability in a given context. We propose three logical steps: \textit{First}, define the main contextual factors, such as who the audience of the explanation is, the operational context, the level of harm that the system could cause, and the legal/regulatory framework. This step will help characterize the operational and legal needs for explanation, and the corresponding social benefits. \textit{Second}, examine the technical tools available, including post hoc approaches (input perturbation, saliency maps...) and hybrid AI approaches. \textit{Third}, as function of the first two steps, choose the right levels of global and local explanation outputs, taking into the account the costs involved. We identify seven kinds of costs and emphasize that explanations are socially useful only when total social benefits exceed costs.\end{abstract}


\pagebreak

\section{Contextual factors drive explainability}\label{sec1}

The recent enthusiasm for artificial intelligence (AI) is due principally to advances in deep learning, and particularly the use of (deep) neural networks. One of the key characteristics of neural networks is that they automatically create and learn their own internal rules from data, without being explicitly programmed. The internal parameters are virtually incomprehensible for humans.\footnote{Neural networks are both highly accurate and  opaque. A highly accurate image classifier can still make mistakes, and the mistakes when they occur will appear shockingly stupid to a human \citep{Dessalles2019}. For example, an image classifier that has been taught to distinguish dogs from wolves may be fooled into thinking that a dog is a wolf simply because the image contains snow \citep{Selbst2018}.}

\medskip

Machine learning is not the only AI technology available. Symbolic AI has been around for decades and includes several discrete branches. Symbolic AI approaches incorporate symbols, logical rules, graphs and/or knowledge bases that reflect human thought. But the various symbolic AI approaches -- called GOFAI, “good old fashioned AI” \citep{Haugeland1985} -- had difficulty dealing with large amounts of real-world data. Hybrid AI approaches seek to capture the best of symbolic and machine learning approaches.

\medskip

Whether symbolic or based on machine learning, any AI system can become so complex as to be inscrutable.\footnote{Inscrutability means that “the rules that govern decision-making are so complex, numerous, and interdependent that they defy practical inspection and resist comprehension” \citep{Selbst2018}.} And inscrutable algorithms cannot generally be deployed in safety-critical applications such as autonomous vehicles, autonomous weapons, medical devices, or industrial maintenance. The ability to explain the inner workings of the algorithm -- to system designers, users, regulators, and citizens -- is critical to achieve trust and accountability. Explainability of algorithms has therefore become a major field of multidisciplinary research, involving data science, statistics, applied mathematics, computer science, sociology, economics and law.  

\medskip

While there is general agreement on the need for explainability \citep{OECD2019a}, policy documents lack clarity on what explainability means and how it should be applied in different contexts. The term ‘explainability’ is vague - it means many things in many contexts. Part of the complexity is due to the fact that explainability comes in many forms, and serves many purposes. Explainability can help users of Google’s search engine understand the PageRank system, or help investigators understand why an autonomous vehicle crashed; it can help doctors learn to trust AI-based diagnostic tools and understand their limitations; it can help detect hidden discrimination in credit scoring algorithms, or improper use of personal data. Explainability can help an individual understand why he or she was singled out for a tax audit or denied admittance to a university; it can reveal statistical bias, such as when a facial recognition algorithm has been trained mostly on faces with white skin. Explainability can permit individuals to exercise their constitutional right to contest government decisions. There are almost as many forms of explainability as there are AI-related problems that need to be understood and solved. For that reason, explainability can never be separated from the underlying AI-related harm or concern (fundamental rights, usability, safety) that explainability is seeking to address.  

\medskip

The purpose of this paper is to unpack the concept of explainability and place it in the broader topic of AI ethics and accountability. We argue that to develop a safe and ethical AI by design, i.e. an AI that respects a set of legal, ethical and safety rules that can be adapted to context, a multidisciplinary approach to explainability is required, and that the approach must use context as the starting point.  

\medskip

Existing works on explainable AI focus on the computer science angle \citep{lime}, or on the legal and policy angle \citep{Selbst2018}. The originality of this paper is to integrate technical, legal and economic approaches into a single methodology for reaching the optimal level of explainability. The technical dimension helps us understand what explanations are possible and what the tradeoffs are between explanability and algorithmic performance. As noted by the European Commission, the "manner in which the information is to be provided should be tailored to the particular context" \citep[p. 20]{EuropComm2020}. Explainability is context-dependent, and context depends on law, economics and sociology. Integrating all these dimensions is necessary to reach flexible and context-specific explainability.

\medskip

\begin{figure}[H]
\centering
\includegraphics[scale=0.8]{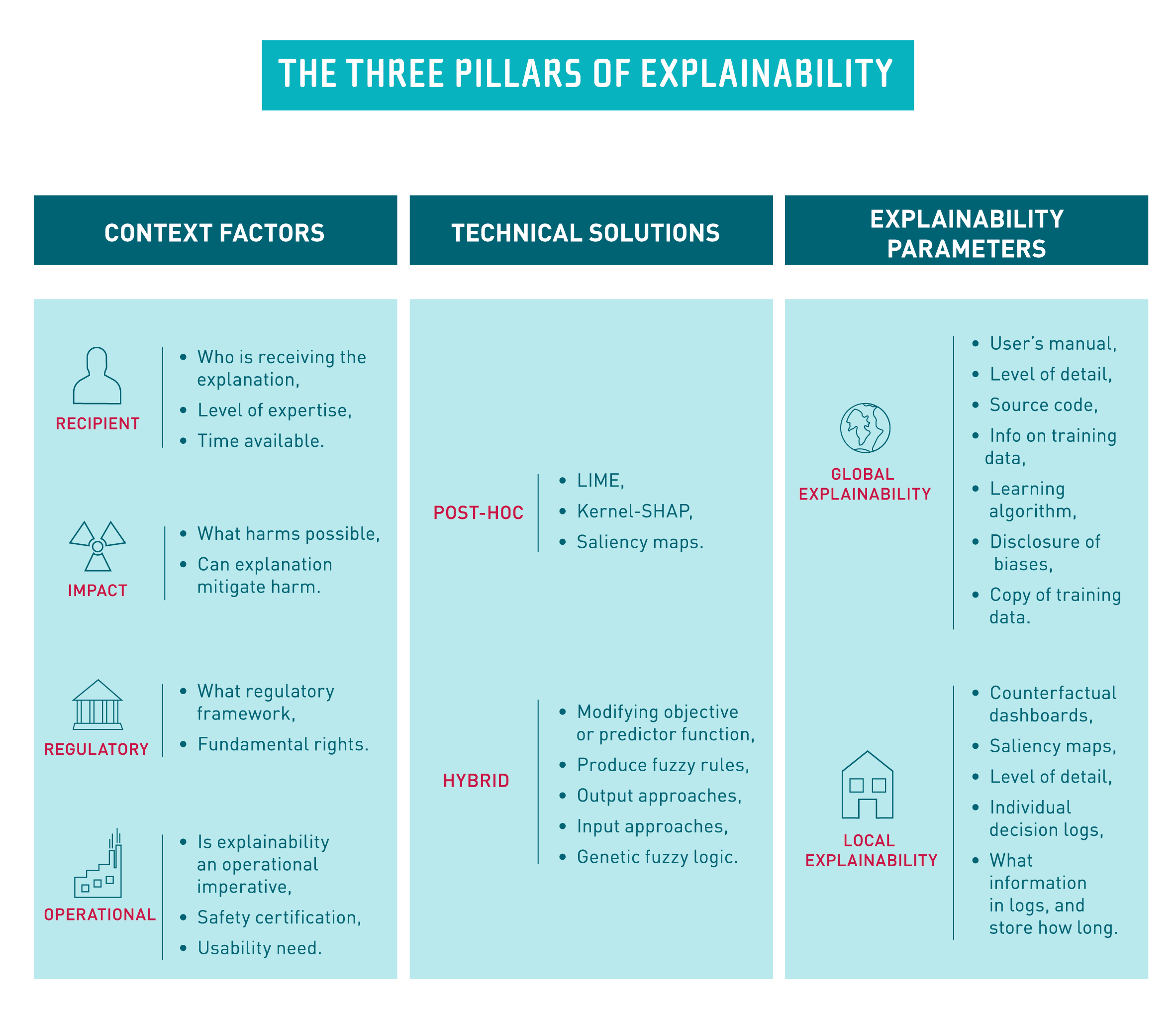}
\caption{Presentation of the three kinds of criteria relevant to algorithmic explanations}\label{VECTO_01}
\end{figure}

Our approach is illustrated in Figure \ref{VECTO_01} and can be summarized in three steps: \textit{First}, define the main contextual factors, such as who is the audience/recipient of the explanation, the operational context, the level of harm that the system could cause, and the legal/regulatory framework. \textit{Second}, examine the technical tools available, including post hoc approaches (input perturbation, saliency maps) and hybrid AI approaches. \textit{Third}, as function of the first two steps, choose the right level and form of global and local explanation outputs, taking into the account the costs involved, including the costs of creating and storing decision logs. The form of output will depend in part on a cost-benefit analysis.

The remainder of the paper is organized in six sections. In Section \ref{sec2}, we provide a brief history of explainability and provide definitions of explainability and related terms such as interpretability, transparency, traceability, auditability and accountability. In Section \ref{sec3}, we review the technical solutions used to provide explanations in different machine learning models and also investigate how to incorporate explainability by design into the model via hybrid approaches. In Section \ref{sec4} we examine the regulatory environment for explainability, and in particular when the law does, and does not, impose explanations. The law will often impose the principle of explainability without defining the details. The exact form and level of explanation will depend in part on the costs and benefits of providing explanations in the relevant context. Section \ref{sec5} examines the economics of explanations, proposing a taxonomy of explainability costs. Sections \ref{sec4} and \ref{sec5} will lead to a better understanding of the context, which in turn dictates what kind of explanation is needed and when. Section \ref{sec6} links explainability to quality control of AI systems, and proposes a governance framework, including an AI impact assessment and review by an independent body. Finally Section \ref{concl} concludes by arguing that flexible and context-specific explainability, including our three step methodology (context factors, technical tools and explainability output choices) will need to be part of the AI impact assessments that are likely to become mandatory for high-risk AI systems.  

\section{Explainability: History, definitions and related concepts}\label{sec2}

Explainability of algorithms is a relatively old concept. We propose in this section a brief overview of the history of the concept, but also a discussion on how the concept of explainability relates to transparency and accountability.

\subsection{A short history of explainability}

The current interest in AI explainability leads us to forget that the subject is at least three decades old  \citep{Selbst2018}. The explainability of artificial intelligence was a vibrant field of research as early as the 1970s \citep{Shortliffe1975}. Many of the AI systems at the time were called "expert systems," bringing together a series of rules and knowledge bases created by humans. The human-created rules became so complex that they defied easy comprehension by users.  This led  computer scientists to study methods to make systems more comprehensible for users \citep{clancey1983epistemology}. In 1991 Swartout et al. developed an "Explainable Expert Systems" (EES) framework in the context of a DARPA program \citep{swartout1991explanations}.  The emphasis was then on symbolic AI. 

\medskip

Multi-layered neural networks have been widely deployed in recent years thanks to the availability of massive data and processing power. Nevertheless neural networks have been studied for several decades. In 1993 Gerald Peterson wrote that neural networks will not be used in critical applications unless people are convinced they can be trusted to perform reliably, a statement that still holds true today. Peterson added that achieving trust requires an understanding of the "inner workings" of the model \citep{peterson1993foundation}. Consequently even the explainability of neural networks has been recognized as an issue for almost thirty years. The difficulties associated with explaining the functioning of neural networks contributed to the decline in machine-learning research during the so-called AI winters.

\medskip

Research on explainability re-emerged in 2016 for three main reasons: First, the deep learning community needed tools to explain the results of multi-layer neural networks. Second, the 2016 European General Data Protection Regulation\footnote{Regulation (EU) 2016/679 of the European Parliament and of the Council of 27 April 2016 on the protection of natural persons with regard to the processing of personal data and on the free movement of such data, and repealing Directive 95/46/EC.} (GDPR) sparked a new focus on explainability as a legal and human rights requirement. Third, industrial users of AI recognized the need for explainability in building safety-critical systems, including the incorporation of explainability in hybrid AI models that combine machine learning and symbolic AI (see Section \ref{sec3} below). 

\medskip

Much of the work on explainability in the 1990s, as well as the new industrial interest in explainability today, focus on explanations needed to satisfy users’ operational requirements. For example, the customer may require explanations as part of the safety validation and certification process for an AI system, or may ask that the system provide additional information to help the end user (for example, a radiologist) put the system’s decision into a clinical context.  

\medskip

\begin{figure}[H]
\centering
\includegraphics[scale=0.7]{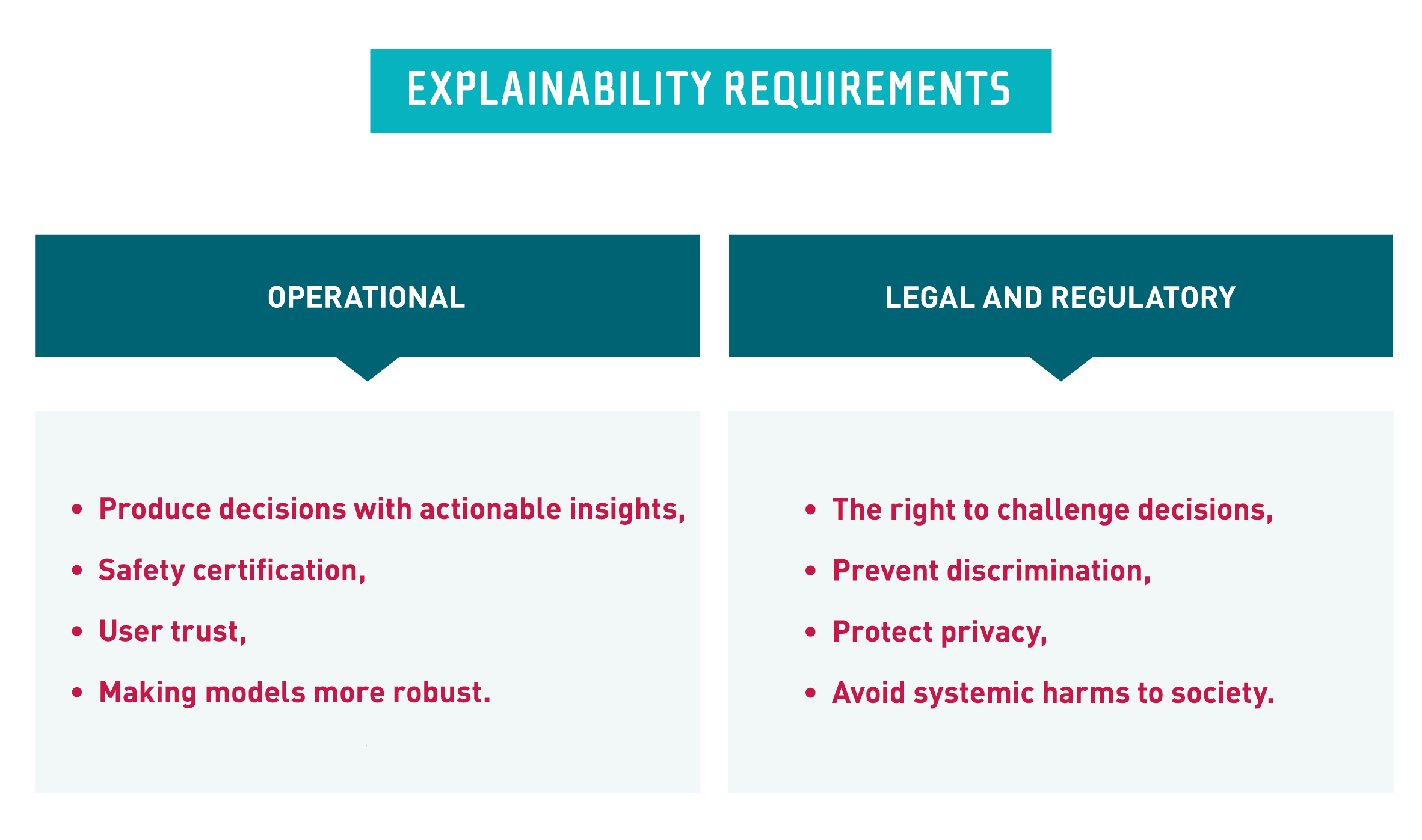}
\caption{Differences between operational and legal/regulatory needs for explanations}\label{VECTO_02}
\end{figure}

These operational requirements for explainability may be required to obtain certifications for safety-critical applications, since the system could not go to market without those certifications. Customers may also insist on explanations in order to make the system more user-friendly and trusted by users. Knowing which factors cause certain outcomes increases the system's utility because the decisions are accompanied by actionable insights, which can be much more valuable than simply having highly-accurate but unexplained predictions \citep{welling2015ml}. Understanding causality can also enhance quality by making models more robust to shifting input domains. Customers increasingly consider explainability as a quality feature for the AI system. These operational requirements illustrated in Figure \ref{VECTO_02} are distinct from regulatory demands for explainability, which we examine in Section \ref{sec4}, but may nevertheless lead to a convergence in the tools used to meet the various requirements.   

\subsection{A taxonomy of definitions: from explainability to accountability}

One of our first challenges is to clarify terminology. The term “explainability” is widely used in academic and policy literature, but does not yet appear in major dictionaries.  

\medskip

Purists of the English language will regret the proliferation of terms including the suffix "-ability": explainability, interpretability, traceability, auditability, accountability. Adding "-ability" to the end of a word transforms its meaning to “the ability, inclination, or suitability for a specified action or condition."\footnote{Unless otherwise indicated, all definitions in this section come from the American Heritage Dictionary of the English Language, Fifth Edition, 2020.} Explainability thereby becomes the \textit{ability, inclination, or suitability} to explain, or provide an explanation. Interpretability becomes the \textit{ability, inclination or suitability} to interpret or provide an interpretation. 

\medskip

To understand the terms explainability and interpretability, we therefore need to focus on the respective meanings of "explain" and "interpret." To explain means "to make plain or comprehensible;" to interpret means "to explain the meaning of," or “to translate from one language to another.” In English,\footnote{In French, the term "interpretable" means "that to which one can give a meaning" (Dictionary of the Academie française, 9th edition), which suggests that in French "interpretabilité" would be linked to the understanding of a particular decision, i.e. "local explainability".} explain and interpret are synonyms. To interpret means to explain, and to explain means to make something comprehensible. Data scientists prefer the word “interpretability,” whereas most policy documents, including the 2019 OECD recommendation on AI \citep{OECD2019b}, now refer to AI “explainability.”  But both terms mean \textit{the ability, inclination or suitability to make plain or comprehensible, or explain the meaning of, an algorithm}.\footnote{Some machine learning practitioners refer to interpretability as the ability to explain the global functioning of the algorithm to an expert, and "explainability" as the ability to explain a particular decision to an end-user.} Implicit in this definition is that the algorithm is made comprehensible to a human \citep{guidotti2018survey}. For purposes of this paper, we will continue to use the term "explainability" and "interpretability" as synonyms.

\medskip

A critical distinction exists between global explainability (or global interpretability), and local explainability (or local interpretability). Global explainability means the ability to explain the functioning of the algorithm in its entirety, whereas local explainability means the ability to explain a particular algorithmic decision \citep{guidotti2018survey}. Local explainability is also known as “post hoc” explainability. The distinction between global and local explainability will be examined in more detail below.

\medskip

Explainability is often associated with transparency, which means “the quality or state of being transparent;” transparent means “open to public scrutiny; not hidden or proprietary.”  In artificial intelligence, transparency generally refers to making information about the inner workings of the algorithm available for scrutiny, including how an AI system is developed, trained and deployed \citep{OECD2019a}. Transparency does not necessarily mean that the underlying information is easily comprehensible to humans. Transparency can simply mean that the information is made available as-is. Additional work may be required to transform the raw information delivered via transparency into something easily comprehensible. Consequently explainability often implies transformation of raw information into something intelligible. 

\medskip

Transparency and explainability are part of a bigger package of measures designed to achieve accountability. “Accountable” means “expected or required to account for one’s actions; answerable.” “Accountability” therefore means the ability, inclination or suitability to make someone answerable for his or her actions. According to the OECD, accountability means the ability to place the onus on the appropriate organizations or individuals for the proper functioning of AI systems \citep{OECD2019a}. 

\medskip

Explainability is an input to accountability, as are traceability and auditability. Traceability means the ability to “ascertain the origin or location of something,” for example the factors that contributed to an algorithmic decision. Traceability will contribute to auditability, because auditors will need to be able to locate information in order to audit (audit means “evaluate thoroughly”) the functioning of the system. Traceability and auditability will generally require the generation of logs, i.e. “a record, as of the performance of a machine or the progress of an undertaking.” The practice of logging consists in the recording of certain program actions in a file either immediately before or immediately after they have taken place \citep{kroll2016accountable}.

\medskip

\begin{figure}[H]
\centering
\includegraphics[scale=0.7]{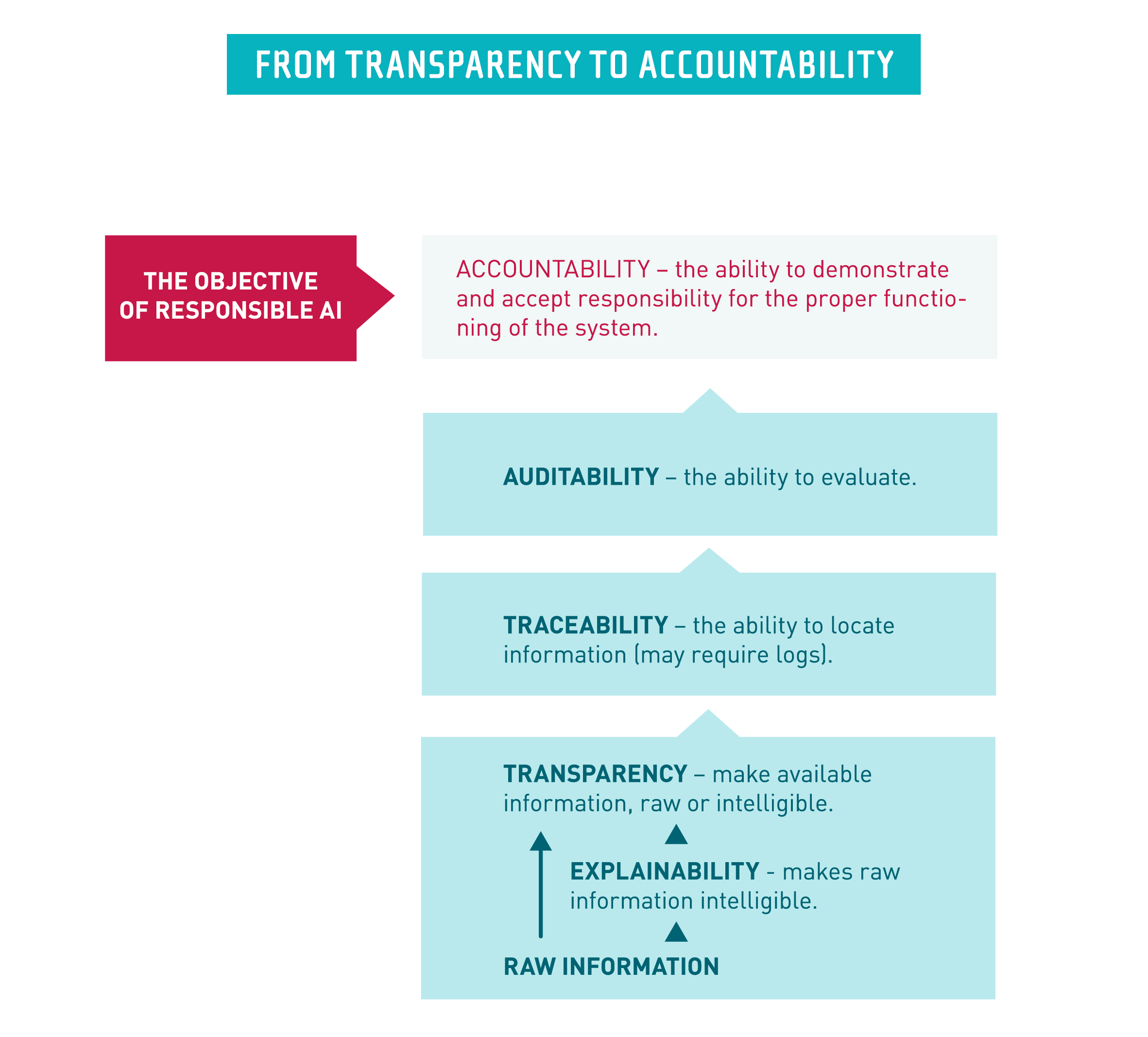}
\caption{Illustration of how explainability relates to transparency and accountability}\label{VECTO_03}
\end{figure}

In summary, explainability and interpretability are part of transparency, and transparency in turn contributes to traceability, auditability, and ultimately accountability, as illustrated in Figure \ref{VECTO_03}.


In our taxonomy, accountability is at the top of the hierarchy, representing the ethical AI objective to which all other elements contribute. Accountability requires a number of desirable AI features to be present, such as respect for principles of human values and fairness, transparency, robustness and safety \citep{OECD2019a}. Human values and fairness require flexible, context-specific explanations, which we believe can only be developed using the three-step methodology presented in this paper: a look at contextual factors, technical tools, and explainability output parameters chosen in part based on a cost-benefit analysis.  

\medskip

Not all authors would agree with our way of presenting these definitions. For example \cite{waltl2018explainable} present transparency and interpretability as subcategories of a bigger concept called explainability, whereas we place explainability as a subset of transparency. Also, some AI practitioners consider interpretability as meaning a global understanding of the system (what we call \textit{global} explainability) and explainability as meaning explanation of a particular decision to a user (what we call \textit{local} explainability). The European Commission \citep{EuropComm2019} presents transparency as a category that includes explainability, traceability and communication to stakeholders. Our approach places transparency further down the value chain than the Commission's approach. These divergences are not important, however, except insofar as they illustrate the lack of consensus on exactly what transparency and explainability mean. 

\medskip

To help sort through the terminology, we have included a glossary in Appendix \ref{appendixA} showing how different terms are defined in various papers dealing with explainability. 

\subsection{Choosing a language}

It is important for systems to provide descriptions in a language suited to the type of explanation needed for the given user.  Explanations oriented around traceability and auditing will likely be more detailed and precise, whereas that same explanation in a decision-support role could overwhelm the user and be counter-productive.

\medskip

Consider, for example, a semi-autonomous vehicle.  A “black box” suitable for diagnosing catastrophic failures would require detailed explanations of all aspects of the vehicle's decision-making processes, such as to identify why the vehicle did not slow for a passing cyclist.  For the driver supervising such a system, the pertinent explanation might simply be that the vehicle does not recognize the presence of the cyclist and that she thus needs to intervene accordingly.  It would be unreasonable to expect such a driver to understand the operational details of the various technological subsystems involved in explaining why the vehicle acted in such a way, just as it would be unreasonable to expect system's engineers to debug the system from a mere indication of the presence of an obstacle.

\medskip

The language of the explanation thus depends on the role and expertise of the relevant audience targeted by the explanation, as well as whether “global” or “local” explanations are required in a particular context.  The following sections explore these latter concepts more fully.

\subsection{Global explanations provide information about the overall functioning of the algorithm}

Global explanations are meant to give an understanding of the overall functioning of the algorithm. Global explanations can take many forms, going from the communication of source code and training data, to simple overviews of the functioning of the algorithm, such as Google’s explanation of how its search algorithms work.\footnote{Google’s explanations are available \href{https://www.google.com/search/howsearchworks/algorithms/}{here}.} The communication of source code, training and testing data would be warranted (and likely required) in a high-stakes lawsuit or investigation relating to an accident. However, in other situations, disclosure of source code may offer little value to global explainability \citep{Selbst2018, ieee2019ethically}.

\medskip

\begin{figure}[H]
\centering
\includegraphics[scale=0.7]{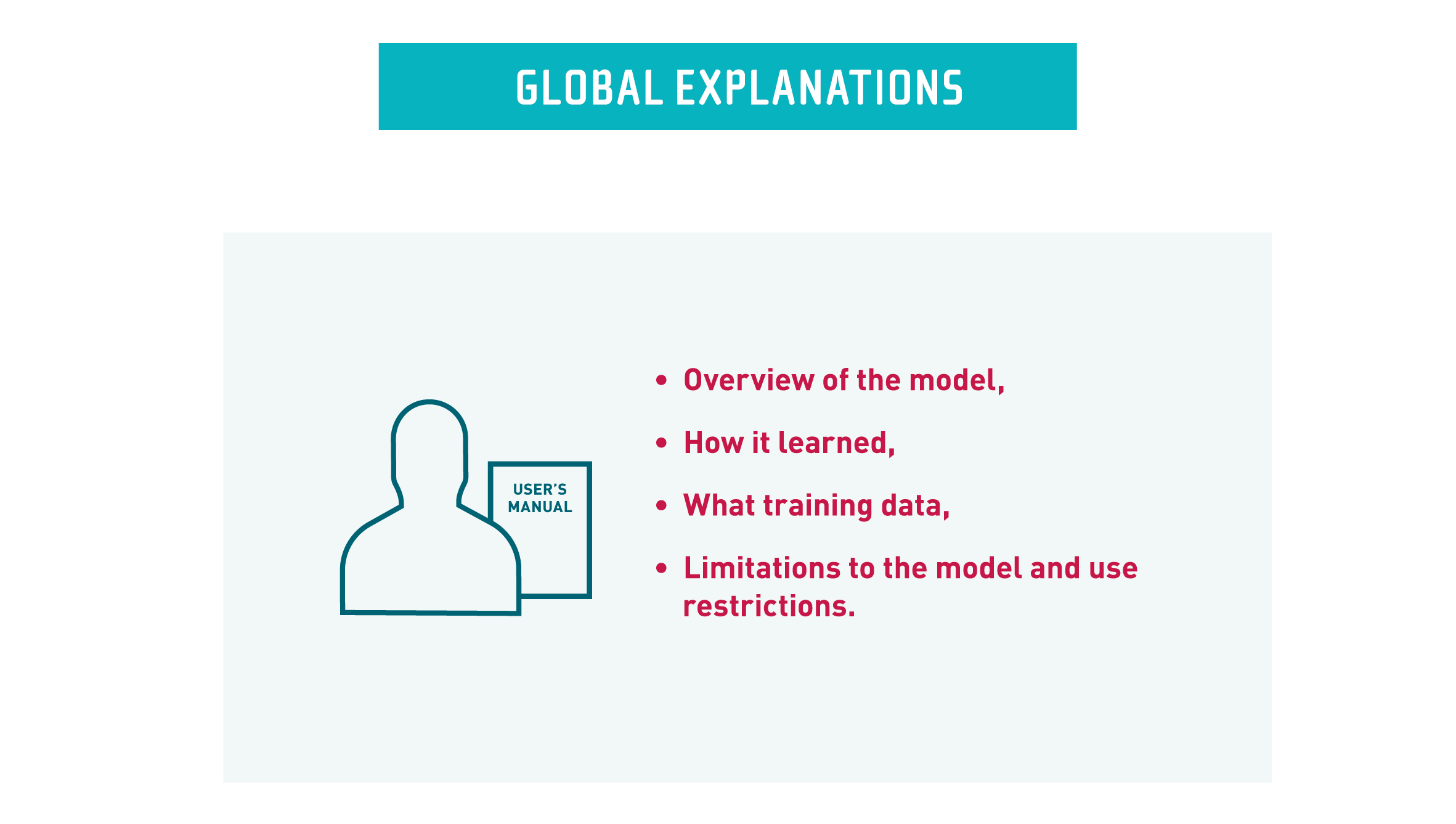}
\caption{Overview of global explanations}\label{VECTO_04}
\end{figure}

For most run-of-the-mill explainability situations, global explainability might be achieved by a descriptive overview of how the algorithm functions (see Figure \ref{VECTO_04}), somewhat like a user’s manual would explain the functioning and use limitations of a car. The IEEE proposes a list of minimum information that might be included in such a user’s manual \citep[p. 245]{ieee2019ethically}:
\begin{itemize}
    \item nontechnical procedural information regarding the employment and development of a given application of autonomous and intelligence systems;
    \item information regarding data involved in the development, training, and operation of the system;
    \item information concerning a system’s effectiveness/performance;
    \item information about the formal models that the system relies on; and
    \item information that serves to explain a system’s general logic or specific outputs. 
\end{itemize}

For the European Commission, the user's manual approach would include "clear information...as to the AI system's capabilities and limitations, in particular the purpose for which the systems are intended, the conditions under which they can be expected to function as intended and the expected level of accuracy in achieving the specified purpose"\citep[p. 20]{EuropComm2020}. The level of detail of the user’s manual will depend on the level of potential risk of the system. Where information might reveal trade secrets, or be misused to manipulate the system, the level of detail would be reduced so as to avoid potential attacks, gaming or misuse. By contrast, to be meaningful explanations about the formal models relied on and general logic involved would need to go beyond the simple affirmation that the model relies on a "neural network that approximates a complicated function." It may be appropriate, for example, to explain the goals that the model seeks to achieve via the loss function.  

\subsection{Local explanations provide information about a particular decision}

\begin{figure}[H]
\centering
\includegraphics[scale=0.7]{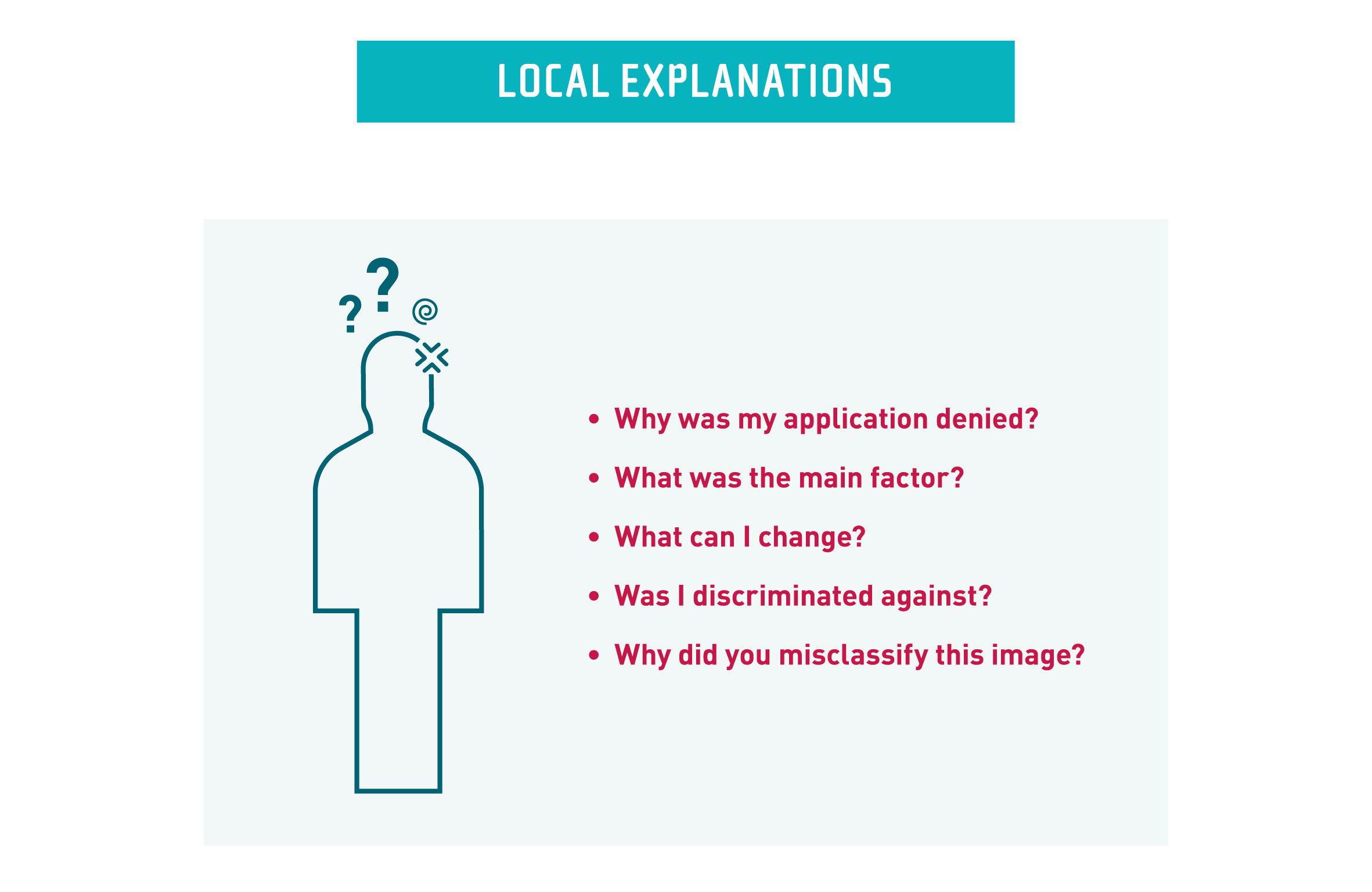}
\caption{Overview of local explanations}\label{VECTO_05}
\end{figure}

Local explanations tell a user why a particular decision was made (see Figure \ref{VECTO_05}). The form of local explanation will vary greatly depending on the audience and the context of the explanation. \cite{doshi2017accountability} posit that a useful local explanation should answer at least one of the following questions: What were the main factors in the decision? Would changing a certain factor have changed the decision? Why did two similar-looking cases get different decisions, or vice versa? The time available to the user is critical. If the user needs to make a quick decision, a simple and fast explanation is required. If time is not a constraint, more detailed explanations would be warranted \cite[p. 6]{guidotti2018survey}. The level of expertise of the person receiving the explanation will also be a factor \citep{ico2019}. An explanation to a data scientist would not contain the same information as an explanation to a radiologist or to a loan applicant.

Counterfactual explanations are one of the best ways to provide useful information to users affected by a decision \citep{wachter2017counterfactual}. Counterfactual explanations focus on “the smallest change to the world that can be made to obtain a desirable outcome” \citep[p. 845]{wachter2017counterfactual}. Allowing users to experiment with different input features to assess their impact on outputs can help users develop trust and an understanding both of a particular decision (local explainability) but also of the algorithm as a whole (global explainability).  Local counterfactual explanations are particularly helpful when they relate to features that a user can control, as opposed to features outside of the user’s control \citep{Selbst2018, ico2019}. For example, a counterfactual explanation demonstrating how different income levels might impact a loan application would be more helpful to the user than a counterfactual explanation showing what would happen if the loan applicant lived in a different country. The ICO refers to individuals’ ability to “learn and change” based on counterfactual explanations \citep{ico2019}. Nevertheless, some counterfactual explanations relating to elements that a user cannot control (for example "What would happen if I were male?") will be helpful to reveal potential discrimination. 

\section{Technical solutions: towards explainability by design}\label{sec3}

We review in this part several technical solutions described in the literature which provide explanations in different machine learning models, and also explore approaches to incorporate explainability by design into AI models.

\subsection{Distinguishing the learning algorithm from the trained algorithm}

\begin{figure}[H]
\centering
\includegraphics[scale=0.7]{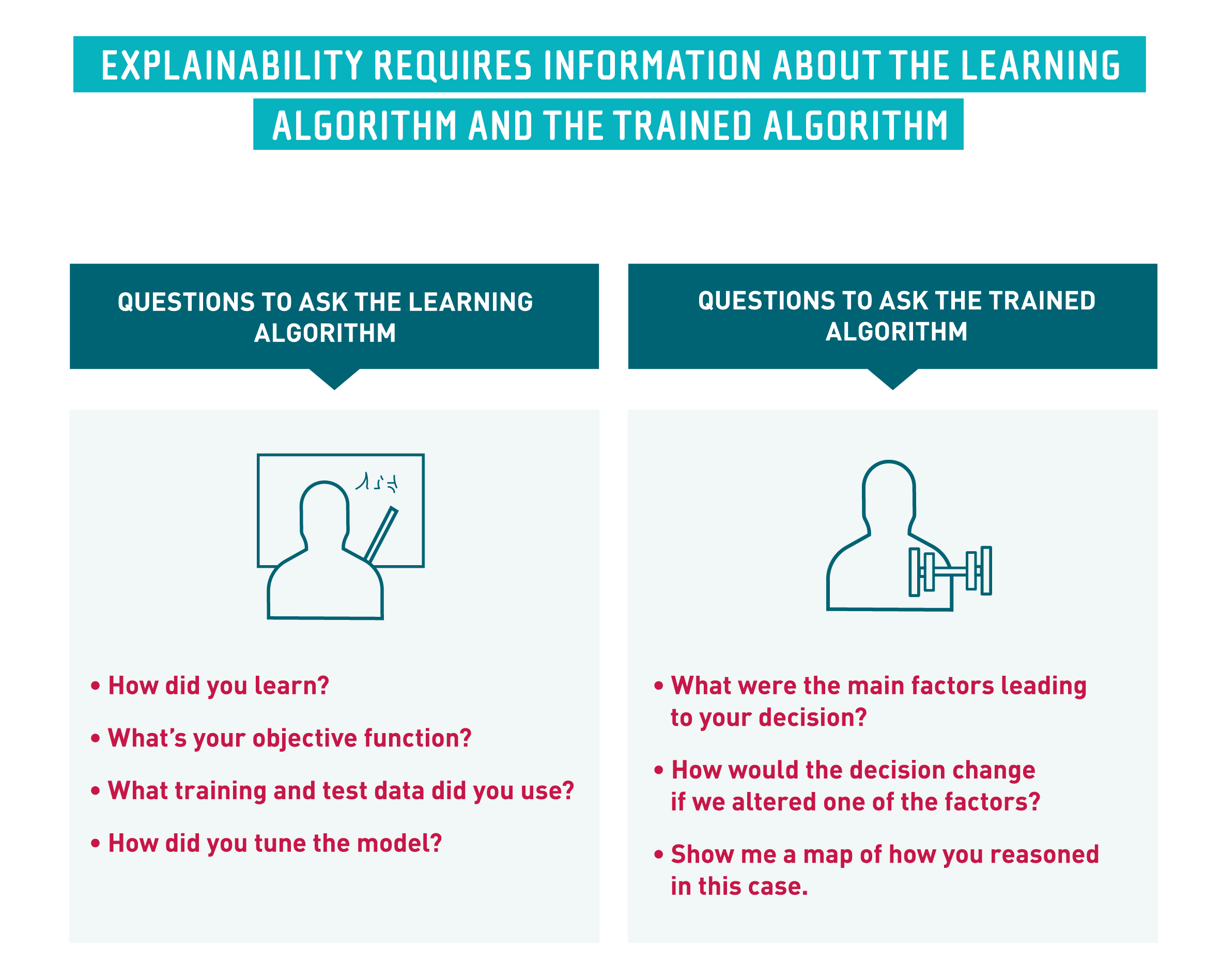}
\caption{Explanations relating to the learning algorithm and to the trained algorithm}\label{VECTO_06}
\end{figure}

Two aspects of machine learning models may require explanation: (i) the learning phase, including the learning algorithm and training data that were used to build the model, and (ii) the operation (or inference) phase in which the trained model makes predictions using real-world data (see Figure \ref{VECTO_06}). Increasingly, models continue to learn after they are put into operation, which means that the learning phase may be repeated numerous times to update the trained model so that its predictions become more and more accurate. 

It is important to consider the learning and operation phases separately. An explanation relating to the \textit{operation phase} will answer questions like: “What were the main factors that led you to identify this as a suspicious transaction?” An explanation relating to the \textit{learning phase} will answer questions like: “How did the model learn to attribute weights to these seemingly unimportant features?”

\medskip

To achieve an overall understanding of how a model functions (global explainability), it may be necessary to understand the main characteristics of the learning algorithm, particularly the goal (the objective function) that the algorithm is seeking to achieve. An explanation of the learning phase would answer the question: “What is the objective function that the learning algorithm seeks to maximize or minimize?” The actual method used by the learning algorithm itself may not be important to explain, whereas the goal and system of rewards helping the learning algorithm reach the goal may be extremely important. A well-known video\footnote{The video is available \href{https://www.youtube.com/watch?v=Q70uIPJW3Gk}{here}.}  showing how Deep Mind learned to play a simple Atari video game shows that the learning process is usually based on trial and error, with little resemblance to the typical learning processes used by humans. Because computers can make millions of iterations – many more than humans can – the trial and error method works well, and exactly how the algorithm learns from its errors (e.g. stochastic gradient descent) may not be important for a user or regulator to understand, whereas understanding the mathematical function that the algorithm is seeking to maximize (or minimize) will be of key importance. Another question relating to the learning phase may relate to the origin and quality of the training data, because poor training data will affect the quality of the trained algorithm. 

\medskip

Explanations about the trained algorithm will help achieve local explainability, i.e. how a particular decision was arrived at.  

\subsection{Different AI models and their explainability}

\subsubsection{Kinds of models}

The world of artificial intelligence can roughly be divided into machine learning (of which deep learning is a part), and symbolic AI.

\medskip

Machine learning can be roughly split into three main task formulations: 
\begin{itemize}
    \item supervised learning, where the output is given for each observation of the training sample;
    \item unsupervised learning, where no output is given, the structure of the data is (entirely or to a certain extent) unknown and should be explored, by categorizing data into meaningful groups;
    \item reinforcement learning, where the environment is explored by actions and awards for the actions, the aim being to develop an optimal (in the sense of the award) strategy.
\end{itemize}

There exist many variations of the foregoing three categories, e.g. weakly supervised learning. Supervised learning is currently more developed than the other two groups of methods, and most of the work on explainability focuses on supervised learning, including deep learning (machine learning through multi-layer neural networks). Even though neural networks are today seen as a most promising classifier, they all share such drawbacks as instability of the classification rule, lack of interpretability, but also an automated methodology for choosing the design, which makes the inner design particularly difficult to understand for a human observer. A more complete description of machine learning methods appears in Appendix \ref{appendixB}.

\medskip

Symbolic AI consists of various approaches to artificial intelligence known as GOFAI - “good old fashioned AI.” Many forms of symbolic AI exist, such as various forms of knowledge representation and reasoning, often based on logics, to solve typical AI problems such as satisfiability, revision, merging, abductive reasoning, and various forms of symbolic learning, including decision trees, association rules, and formal concept analysis. The best known incarnations of symbolic AI rely on decision trees, rule lists, or rule sets although many other forms of symbolic AI exist. Given enough time, a human can understand the logical steps involved in an algorithmic decision based on a decision tree or rule list. However, if the decision tree has hundreds or thousands of branches, explaining a decision to a user will require shortcuts and explanation methods adapted to the particular user. Explaining expert system decisions was the field of research in the 1970s and 80s \citep{Shortliffe1975, clancey1983epistemology}. 

\medskip

Most of the literature on explainable AI focuses on supervised learning. Therefore the sections below focus principally on this technical angle to explainability. 

\subsection{Post hoc approaches for local explainability}

Most current approaches to explain reasons for local predictions of complex models such as deep neural networks and random forests opt for a post hoc analysis and perform attribution over input features. Post hoc approaches provide local explainability for a particular decision. Post hoc approaches can be primarily classified into two groups: (i) approaches that perturb the input to create multiple input output pairs and then fit a simple model to explain the predictions locally, or (ii) approaches based on saliency maps which assign importance scores/attribution to each input feature. We discuss each class of methods and their representative works in more detail below, and also discuss some approaches that tackle the problem differently.

\subsubsection{Perturbation based methods}

This group of methods treat the underlying predictor function as a black box and attempt to explain its prediction by fitting a simple explanor model to multiple perturbed versions of input-output samples. The explaining model does not open the black box, but simply attempts to imitate its behavior by observing how the black box responds to different input perturbations, including occlusion of certain features. The approach proposed by \cite{lime} is one of the earliest proposed systems, representative of this class. For a given sample and a prediction model $f$, LIME (which stands for Local Interpretable Machine-Agnostic Explanations) optimizes a loss function over the class of simple models such as sparse linear models or decision trees to explain predictions for the particular sample. The loss function consists of a term that promotes the model to utilize as few features as possible and a second term promoting fidelity of the explaining model $g$ to $f$ in the neighborhood of the given sample. The explaining model $g$ utilizes a simplified data representation for all the samples, to produce explanations that are understandable to humans. This representation consists of relatively higher level of features compared to the original input processed by $f$.

\medskip

The Kernel-SHAP approach \citep{shap} is an improvement over LIME. It also operates in the same framework. However it imposes certain desirable properties on the explaining function which leads to a unique optimal in the class of linear additive functions that is strongly linked to estimation of classic shapley values.  

\medskip

Although the primary focus of methods from this class is on explaining classification and regression problems there is also an interest in the community to explain predictions of sequence to sequence models. This problem is much harder to fit in the above framework as it is a lot more complicated to perturb a sample and to generate human-understandable explanations. Work by \cite{socrat} is the first attempt in this direction. They propose a novel automatic perturbation algorithm for structured input-output data based on variational autoencoders and produce explanations as bipartite subgraphs consisting of input and output tokens that are causally related. 

\subsubsection{Saliency map based methods}

Methods from this class attempt to identify the set of input features that appear salient to the predictor function by performing sensitivity analysis. In other words they perform attribution with regard to input features by evaluating the sensitivity of output due to each feature. Most of the work in this group has focused on processing of images and explaining predictions of neural networks.

\medskip

\cite{saliency} first proposed the use of saliency maps for explaining convolutional neural network (CNN) decisions on image classification. Their saliency map computation primarily relied on computing the gradient of output with regard to input image pixels. \cite{springenberg2014striving} proposed a modified algorithm for performing gradient backpropagation in CNN’s called guided backpropagation. \cite{gradcam} further improved upon the output of guided backpropagation. They proposed to combine activation maps of the last convolutional layer of the CNN according to their average gradient with regard to output, referred to as class activation maps. These maps are further integrated with guided backpropagation to yield improved results. SmoothGrad \citep{smoothgrad} generates multiple noisy versions of input and averages each of their attribution maps to obtain a sharper and more robust saliency map. \cite{ig} propose axioms for attribution methods and propose their own attribution algorithm for neural networks based on integrating gradients along a path.  Some researchers have questioned the reliability of saliency maps \citep{kindermans2017unreliability}.  

\subsubsection{Other approaches}

Some of the more recent works on explaining black-box models have attempted to tackle the problem differently. \cite{vibi} propose to solve explainability using information-theoretic principles. More specifically, they formulate an optimization problem guided by the information bottleneck principle \citep{tishby2000information} to perform attribution over the input features and select a subset of them as their explanation. Work by \cite{kuhn2019application} adopts a combinatorial approach and presents the general classification problem as a fault localization problem to determine explicit rules for explaining a decision. \cite{kim17tcav} propose a method to quantify sensitivity of a neural networks prediction to high-level concepts, thus quantifying interpretability in terms of human-friendly concepts. They introduce the idea of a concept activation vector which represents the concept in terms of activations of different layers of the network. The sensitivity to a concept is computed by measuring the sensitivity of prediction to perturbations in the direction of the concept activation vector.

\subsubsection{Comparing the two classes}

Perturbation-based methods and saliency map based methods each offer their own sets of benefits and drawbacks. The choice between the two often depends on the context of the problem and its requirements.

\medskip

Perturbation models, the first class, are more flexible. They treat the predicting function as a black box and thus can be applied in a wider variety of settings. Moreover they also generally yield higher level explanations compared to saliency map based approaches and can even provide explanations in terms of some predetermined features. However these advantages also carry drawbacks. Use of simplified data representation also makes these models highly reliant on the quality of these features. They also are much more likely to fail to capture what the original model learned, because they process the simplified representation as opposed to the original input. They also cannot access any internal elements of the predicting model. These approaches to explanation make an educated guess at what is really happening inside the black box model, but there is no guarantee that the explanation actually matches the reality of the model.

\subsection{Explainability by design and hybrid AI: towards incorporating explainability in the predictor model}

As noted before, an important set of methods infer explanations via a post hoc analysis locally. This is not ideal, as the explaining models often do not process the input like the predictor function itself, which can lead them to miss the representations and concepts the predictor function learned. The explanation is only an approximation of the underlying model itself. 

\begin{figure}[H]
\centering
\includegraphics[scale=0.7]{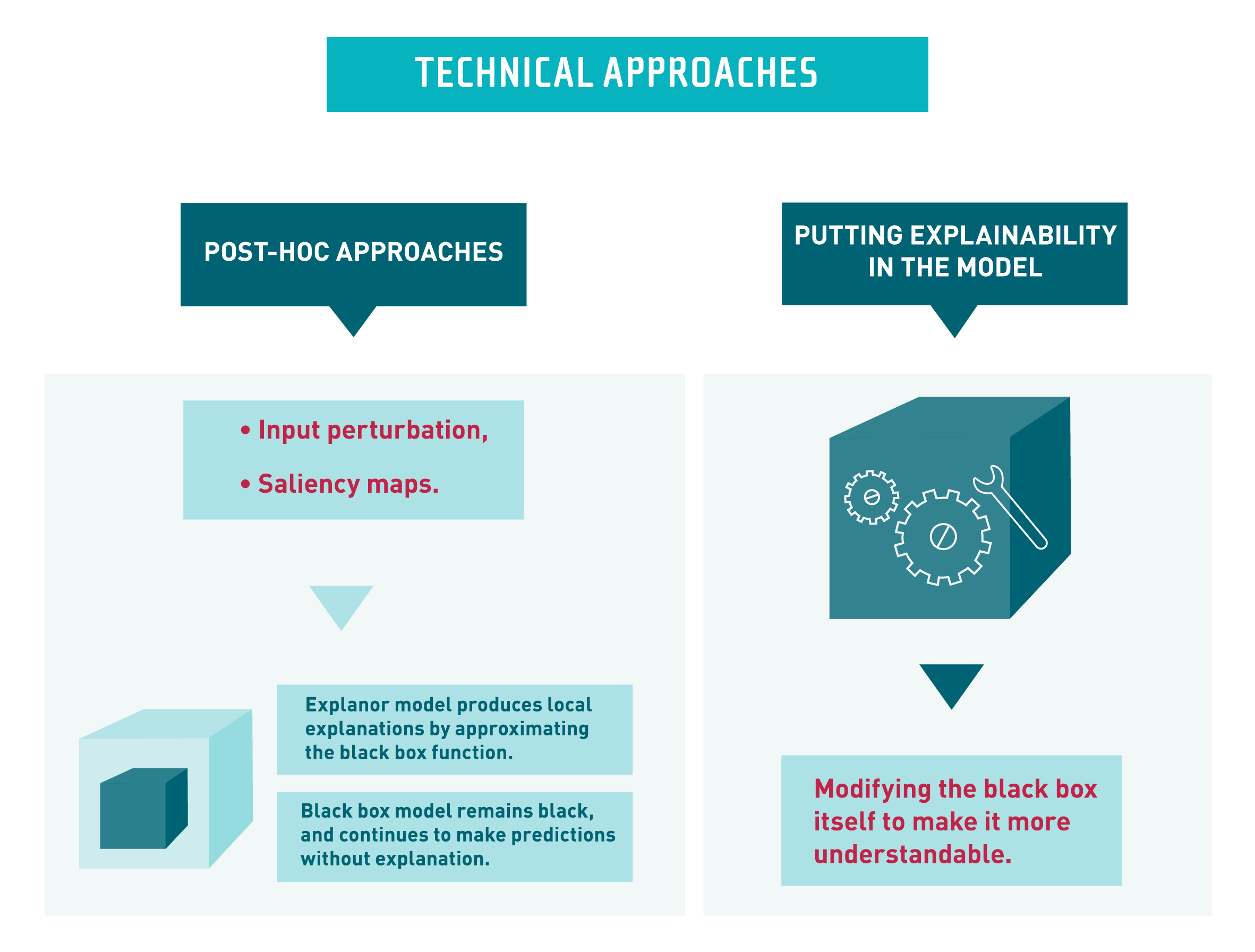}
\caption{Overview of technical approaches to explainability}\label{VECTO_07}
\end{figure}

Currently, methods are gradually moving towards alleviating these problems by modifying the structure of predictor function or changing its training objective. This attempt to reach “explainability by design” is not new, and is part of a more general trend toward "hybrid" AI. Figure \ref{VECTO_07} summarizes the main differences between post hoc approaches and explainability by design. We discuss below recent works for both types of explainability by design approaches (modifying the objective and modifying the predictor function) and connect them to earlier works developed for networks with one hidden layer.
 
\subsubsection{Modifying the objective}

\cite{game} propose a game-theoretic approach wherein the predictor and local explanor, corresponding to each sample, are part of a game. During training, each local explanor is supposed to minimize its deviation from the predictor in a neighborhood of data around its sample, whereas the predictor is supposed to maximize its accuracy while simultaneously minimizing its aggregated deviation with all the local explanors. \cite{icnn} proposed a modification in the loss function of training convolutional neural networks where the loss function for the last convolutional layers apart from maximizing accuracy also promotes each filter to activate only in a single spatial location and for a single class. \cite{roll} propose changes in loss function that stabilizes the gradients in the input space for a particular class of convolutional neural networks and makes them more robust and reliable. These approaches reward the model for adopting internal methods that are explainable. 
\subsubsection{Modifying predictor function}

Contextual Explanation Networks \citep{cen} incorporate the simplified data representation and generation of explanations in the predictor function itself. To avoid losing richness of information in the original input they propose to process the original input through a neural network to produce a weight vector but force the model to make the final predictions by combining the weight vector with a simplified representation. For classification problems they combine the weight vector with a linear model, and for structured prediction tasks they combine the weight vector with a simple graphical model (e.g. Markov chain). 

\medskip

Self Explanatory Neural Networks \citep{senn} generalize the linear model formulation to model weights and features as a function of the given instance. The instance-wise weights are computed  using neural networks. They also compute the features via a neural network but impose many conditions on it such as sparsity and input reconstruction so as to promote the encoding of high level concepts in the features. The individual features are also represented by samples from the data set, to provide insights into the concepts encoded by those features.

\subsubsection{Hybrid AI}

Hybrid AI consists in combining several AI techniques in order to get the best from each method. Hybrid approaches can help enhance reliability, explainability and/or safety. Focusing solely on neural network related approaches, a number of hybrid approaches exist.  One approach uses knowledge to pre-process the inputs to the machine learning model, making the inputs more meaningful and/or better structured for the machine learning model. Another approach focuses instead on the outputs. For example, models flowing from symbolic AI, including rules developed by subject matter domain experts, can constrain outputs from the ML model, acting as a form of output filter \citep{Thomas2019}. A model might also provide a confidence or reliability score simultaneously with the decision, the confidence or reliability score being generated using traditional statistical methods. Two decades ago the explainability of neural networks was addressed by focusing on one-hidden-layered neural architectures equivalent to fuzzy logic rule bases. Interested by the explainability of Radial Basis Function networks, \cite{d1994rule} introduced a learning algorithm that minimizes a modified objective function that reflects the desirable properties of a logic rule base: correctness, consistency and completeness. \cite{Mascarilla1997} proposed  a neural network able to simultaneously produce fuzzy rules, with their linguistic approximation in the language of the user (hence with direct interpretability capabilities), and a final classification.


\medskip

\medskip

In another approach, a hybrid neural network would first be structured to reflect a set of rules created by domain experts, using a “rules to network” algorithm. The network structure at that point represents the symbolic knowledge of the domain. Then the network would be trained on training data, with the objective of refining or supplementing the initial rules based on observed correlations in the training set. The last step is to extract the rules from the network \citep{kurd2003}. 

\medskip

Neural networks can be used to help solve symbolic AI problems, for example so-called satisfiability problems involving many formulas whose compatibility must be tested \citep{selsam2018learning}. Another example consists in integrating logical reasoning with deep learning, for example TensorLog which considers probabilistic first order logic \citep{cohen2017tensorlog}. 

\medskip

\begin{figure}[H]
\centering
\includegraphics[scale=0.7]{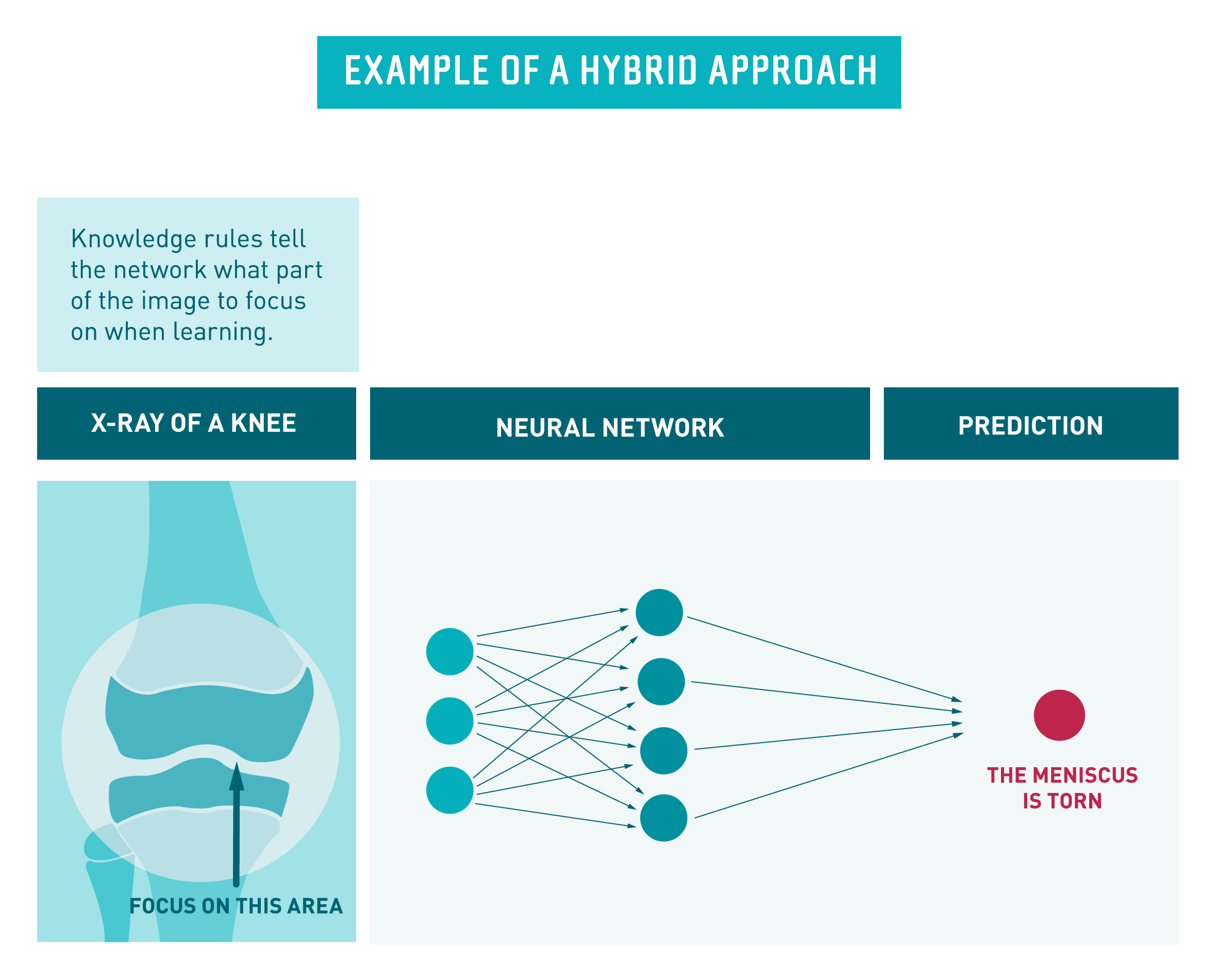}
\caption{Example of showing a neural network which part of the image is important}\label{VECTO_10}
\end{figure}

Hybrid AI approaches can also consist in inserting constraints into the neural network, for example structural information on the location of objects in an image: a chair generally sits on the floor; the spleen is located below the heart in a human body; drivers are located inside cars. Figure \ref{VECTO_10} illustrates how a neural network can be coached, via medical knowledge, on the location of the most important part of a medical image.

Similar to the approach illustrated in Figure \ref{VECTO_10}, \cite{chen2019looks} propose a deep network architecture called “prototypical part network” that focuses on parts of the image that are the most important compared to identified prototypes, and provide the most relevant explanations. The classifications provide explanations based on comparison with learned prototype images. 

\medskip

A hybrid method called “genetic fuzzy logic” consists in embedding in the network a large collection of rules -- a form of dictionary of all permitted actions the system can take -- and then use training data to teach the algorithm how to select the optimal set of permitted rules in a given situation. This approach has been proposed for air combat contexts \citep{ernest2016genetic}, but can be adapted to other contexts such as surgery. It facilitates explainability because in any given situation, the only possible outcomes are those appearing in a pre-approved rule dictionary. Knowledge-based constraints may also be integrated into a neural network to assist in transfer learning, for example when a neural network is trained on medical images of adults, and must be retrained using images of children.

\medskip

Another interesting example of hybrid AI consists in using machine learning methods to learn parameters of rules or other types of knowledge/information models. For example, a rule may state that a robot should not get “too close” to a human; interpreting what “too close” means in a given context is then learned from examples. Because many legal and ethical concepts are context-dependent, this form of parameter learning can be valuable for creating ethical AI “by design.”

\medskip

\begin{figure}[H]
\centering
\includegraphics[scale=0.7]{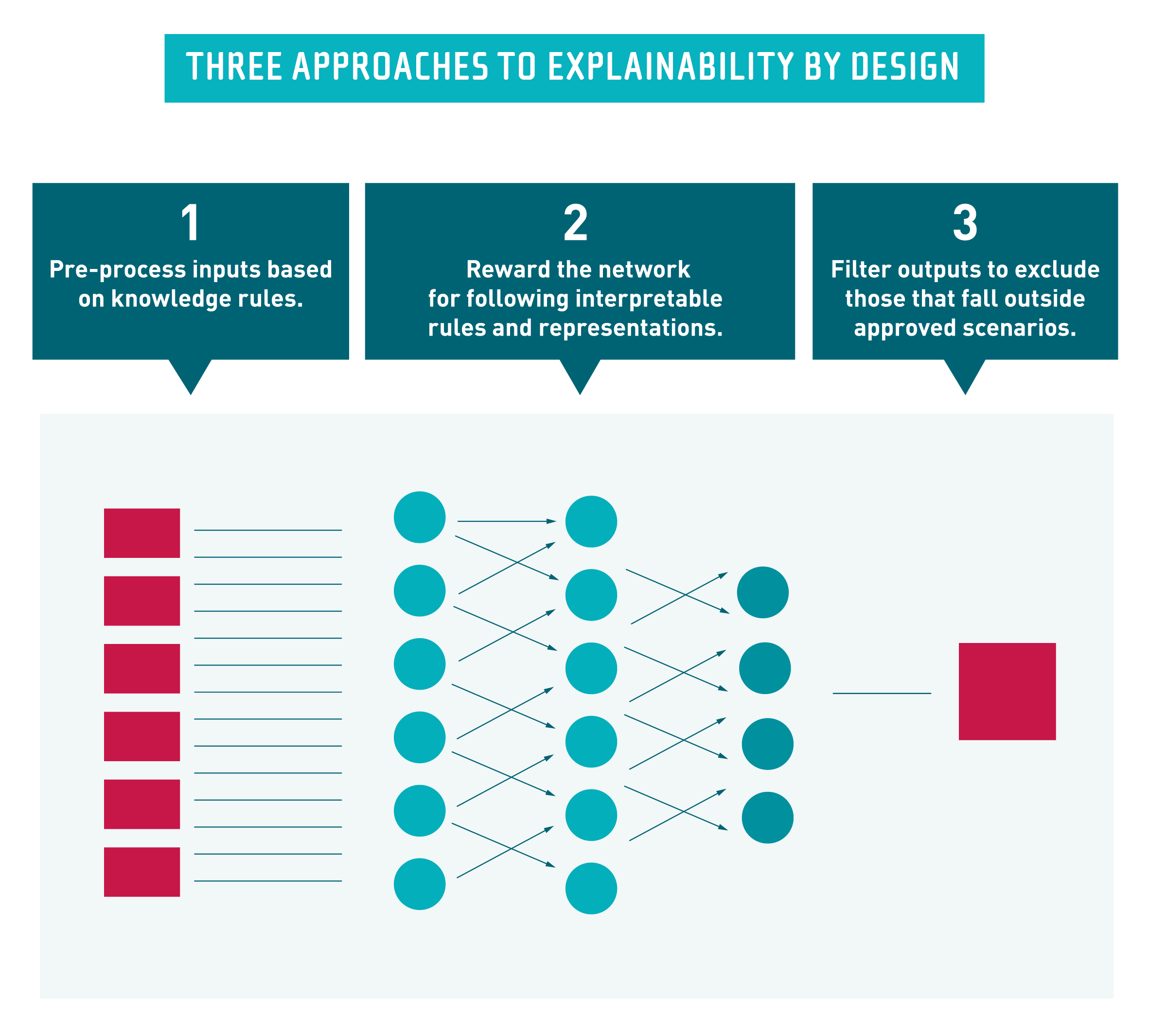}
\caption{Overview of hybrid approaches to explainability}\label{VECTO_08}
\end{figure}

The range of potential hybrid approaches is almost unlimited. The examples above represent only a small selection. Most of the approaches, whether focused on inputs, outputs, or constraints within the model, can contribute to explainability, albeit in different ways. Figure \ref{VECTO_08} illustrates in extremely simplified form the three conceptual approaches discussed above.

\section{Explainability as an ethical and regulatory requirement}\label{sec4}

Having looked at the technical approaches to explainability, we focus in this section on the regulatory environment, and in particular when the law does, and does not, impose explanations. As we will see, the law often imposes the principle of explainability without defining the form and level of detail of the explanation. To fill the gap and define the “right” level of explanation, contextual factors, including a cost-benefit analysis, must be taken into account. We develop in Section \ref{sec5} a taxonomy of the relevant costs and a framework for transforming contextual factors into a cost-benefit analysis. 

\subsection{Regulatory requirements are needed to produce a socially-optimal level of explanation}

Explainability has recently emerged as an ethical and regulatory requirement for AI systems, linked to concerns over hidden discrimination, potential harms to privacy, democracy, and other societal values. The costs of these harms are often not taken into account by the customers and suppliers of AI systems.  A regulatory constraint is therefore needed to address these externalities and produce a socially optimal level of explanation for AI systems.  These regulatory requirements are reflected in the 2019 recommendation of the OECD \citep{OECD2019b} and the European Commission's white paper on AI \citep{EuropComm2020}, and will likely form the basis of new AI regulations worldwide. 

\medskip

The regulatory requirements for explainability are designed to protect several individual and collective interests, including an individual’s right not to be subject to an unfair decision, either by government or by private enterprises, an individual’s right to understand decisions in order to challenge them and/or adapt his or her own behaviour, and an individual’s right to personal autonomy and to protection of his or her personal data \citep{ico2019}. Citizens also have a collective right to know about the existence of algorithms, whether deployed by government or by private enterprises. This requirement is part of the original Fair Information Practices (FIPs)\footnote{Fair Information Practices are a set of five principles that first appeared in the United States Privacy Act of 1974, and later expanded upon in the OECD's 1980 guidelines on privacy, the Council of Europe's 1981 Convention 108 on personal data, and the European Union's 1995 Directive 95/46 on data protection, now replaced by the GDPR.} and forms the basis of privacy regulation in many parts of the world: “there must be no personal-data record-keeping systems whose very existence is secret”  \citep{Selbst2018}.

\medskip

The legal approaches to explanation are different for government decisions and for private sector decisions. Most democratic societies have freedom of information acts or similar government transparency rules that require government decisions, algorithmic or otherwise, to be supported by explanations. The obligation for governments to give explanations has constitutional underpinnings. For private sector decisions, the obligation to give explanations depends on the entity’s market power, or the trust relationship that may exist between the private entity and the individual. We examine below the regulatory environment for public sector decisions, and then for private sector decisions.

\subsection{Algorithms used by governments are subject to higher explainability requirements}

For government decisions, the law generally requires that decisions be explained so that people can challenge the decisions if they disagree. Reasoned decisions are part of the constitutional right of procedural due process in the United States, also reflected in the right to an effective remedy and the right to good administration in the EU Charter of Fundamental Rights. 

\subsubsection{Government use of algorithms triggers a constitutional right to explanation.} 

In an important case in the United States\footnote{\textit{Local 2415 v. Houston Independent School District}, 251 F. Supp. 3d 1168 (S.D. Tex. 2017).}, the court found that an algorithmic scoring system used to rank teachers had to be open to scrutiny in order to permit the affected teachers to verify the accuracy of their score and challenge the decision if they disagree. The court found that without access to "value-added equations, computer source codes, decision rules, and assumptions," teachers could not exercise their constitutionally-protected rights to due process.\footnote{Id., p. 17.} The court said that teachers needed to be able to replicate the algorithmic decision to verify whether there was an error in the score. Without this ability, the teachers’ scores remain "a mysterious ‘black box’, impervious to challenge."\footnote{Id., p. 17.} A court in the Hague, Netherlands, reached a similar conclusion with regard to an algorithm used by the Netherlands government to predict the likelihood of social security fraud.\footnote{\textit{NJCM v. the Netherlands}, District Court of The Hague, Case n° C-09-550982-HA ZA 18-388, February 5, 2020.} The court analyzed the algorithm under the EU test of proportionality, and found that the lack of explanation for the computer-generated risk reports prevented individuals from being in a position to challenge the reports, and prevented the court from verifying the absence of discrimination.  

\medskip

Another important case in the United States involved a judge's use of the COMPAS criminal justice algorithm, which calculates a probability that a particular person will commit another crime if released. The purpose of the system is to help prison and rehabilitation officials determine whether certain prisoners can be put into alternative sentencing programs. However, the system was also used by a judge as a source of information to fix the sentence itself. The affected individual, Mr. Loomis, challenged the decision in court, arguing that the COMPAS system is opaque, that the source code is unavailable, and that the system is racially biased. The Supreme Court of Wisconsin found that the judge’s use of the algorithm did not affect Mr. Loomis’s constitutional rights because the algorithmic score was an insignificant element in the judge’s decision, the judge relying almost exclusively on other factual elements. The United States Supreme Court decided not to hear the case.  Although the Wisconsin Supreme Court decided not to overturn the sentencing judge’s decision, and did not require disclosure of COMPAS source code, the court said that algorithmic tools like COMPAS must be accompanied by a warning statement on the algorithm’s limitation, including information regarding on the population sample used to train the scoring system, and the fact that the population sample may not correspond to the relevant local population. The documentation should disclose that the owners of the algorithm refuse to give access to source code, that studies have shown the system to disproportionately classify minority offenders as having a higher risk of recidivism, and that the system needs to be "renormed for accuracy due to changing populations and subpopulations."\footnote{\textit{State of Wisconsin v. Loomis}, Supreme Court of Wisconsin, n° 2015AP157-CR, July 13, 2016, paragraph 66.} The court  imposed an obligation of global explainability to help users -- in this case judges --  apply appropriate skepticism when using algorithmic tools. The court emphasized that an algorithm may in no case be used to determine the sentence itself, but only as an additional source of information on how the sentence should be served. 

\medskip

For government decision-making, explanations seek to provide information on why the government agency reached a particular outcome (Coglianese and Lehr, 2019, p. 16). However, explanation may also serve to reveal how governments use particular algorithms. In government decision-making, there are two interests at stake: individual due process - an individuals' ability to understand and challenge a decision -- and the interests of citizens collectively to be informed about governmental use of algorithms. 

\subsubsection{For government algorithms, French law requires disclosure of “parameters and their weights”}

The French law relating to the transparency of government algorithms\footnote{ French Code of Relations between the Public and the Administration, articles L. 311-3-1 et seq.} requires disclosure of fairly detailed algorithmic information to users who request it: the degree and manner in which the algorithmic processing contributed to the relevant decision; the data used for the processing and their source; the parameters used and their weights in the individual processing; and the operations effected by the processing. The law requires both global and local explainability, and users must be informed at the outset that the processing involves an algorithm, otherwise the administration's decision is null and void. By requiring disclosure of the parameters used and their weights, the French law makes it difficult for the  government to rely on convolutional neural networks for any individual decision, unless one of the technical explainability tools referred to in Section \ref{sec3} is deployed. 

\medskip

France’s Data Protection Act\footnote{French Data Protection Act 78-17 of January 6, 1978, article 47.} imposes additional requirements: any algorithm used by the government as the sole basis for individual decisions creating legal or other significant effects may not use sensitive data, and the government must ensure that it “masters” the algorithm. According to the French Constitutional Court, “mastering” the algorithm means that the government must be able to "\textit{explain, in detail and in an intelligible form, to the affected individual, the manner in which the algorithm functioned in the individual’s case}."\footnote{French Constitutional Court Decision n° 2018-765 DC of June 12, 2018, paragraph 71.} This excludes, according to the court, the use of an algorithm that “modifies its own rules without the control and validation of the controller."\footnote{Id.} The Constitutional Court's language “in detail and in an intelligible form” now appears in the French Data Protection Act, thereby imposing a high level of local explainability. The use of convolutional neural networks would appear prohibited in the absence of robust explanation tools.

\subsection{Algorithms used in the private sector}

Laws are also emerging on algorithmic transparency in the private sector. Unlike governments, private companies do not have a general duty of transparency and accountability to customers and citizens. A grocery store does not have to explain why it no longer stocks a given brand of potato chips. Imposing a general duty of explanation on all private entities would be disproportionate because in many situations the social costs of explanations would outweigh the benefits. 

\medskip

For private entities, a duty of explanation generally arises when the entity becomes subject to a heightened duty of fairness or loyalty, which can happen when the entity occupies a dominant position under antitrust law, or when it occupies other functions that create market power or a situation of trust or dependency vis à vis users. A number of specific laws impose algorithmic explanations in the private sector. We examine some of them below.

\subsubsection{Europe’s GDPR requires “meaningful information about the logic involved”}

Europe’s General Data Protection Regulation EU 2016/679 (GDPR) creates an obligation of explanation whenever an entity uses an algorithm to make individual decisions that create legal or similar significant effects for an individual, and there is no meaningful human intervention. The GDPR requires that the explanation  include "meaningful information about the logic involved."\footnote{Regulation 2016/679, article 13(2)(f).} The GDPR also provides individuals with the right to obtain “human intervention, to express his or her point of view, and to obtain an explanation of the decision reached after such assessment and to challenge the decision.”\footnote{Regulation 2016/679, recital 71.} The GDPR applies both to private entities and to governments, although as mentioned above, the rules applicable to government algorithms often go further, as is the case in France, by requiring detailed explanations, going beyond the GDPR’s requirement to provide only “the logic involved.”

\subsubsection{Europe’s Platform to Business Regulation requires a “reasoned description” of the “main parameters”}

Europe's Platform to Business Regulation (EU) 2018/1150 represents the most developed efforts to date to put algorithmic explainability into European law. The regulation imposes a duty of explanation on online intermediaries and search engines with regard to ranking algorithms. The language in the regulation shows the difficult balance between competing principles: providing complete information, protecting trade secrets, avoiding giving information that would permit bad faith manipulation of ranking algorithms by third parties, and making explanations easily understandable and useful for users. Among other things, the online intermediary or search engine must provide a “reasoned description” of the “main parameters” affecting ranking on the platform, including the “general criteria, processes, specific signals incorporated into algorithms or other adjustment or demotion mechanisms used in connection with the ranking.”\footnote{Regulation 2018/1150, recital 24.} Online intermediation services and search engines are not required to disclose the detailed functioning of their ranking mechanisms, including algorithms, nor to disclose trade secrets.\footnote{Regulation 2018/1150, recital 27.} However, the description must be based on actual data on the relevance of the ranking parameters used. The explanations must be given in "plain and intelligible language,"\footnote{Regulation 2018/1150, recital 25.} and permit business users to have an "adequate understanding of the functioning of the ranking in the context of their use" of the services.\footnote{Regulation 2018/1150, recital 27.} 

\medskip

The "reasoned description" given by online intermediaries and search engines needs to be useful, i.e. it "should help business users to improve the presentation of their goods and services."\footnote{Regulation 2018/1150, recital 24.} The content of the explanation is tied almost exclusively to its intelligibility and utility by users. Whether an explanation is sufficient to provide an "adequate understanding" in a given situation will no doubt generate disputes. Anticipating this, the regulation empowers the European Commission to publish guidelines to help online platforms understand what they have to do. \footnote{Regulation 2018/1150, article 5(7).}   

\subsubsection{French law on AI-powered medical devices requires traceability}

In the field of health care, a proposed French law on use of algorithmic tools in the diagnosis or treatment of disease would require that doctors inform patients of the existence and modalities of any big data processing operations that contribute to disease prevention, diagnostic, treatment of the patient.\footnote{Proposed Article L 4001-3 of the French Public Health Code, as proposed in Article 11 of the Bioethics Bill adopted by the National Assembly in first reading on October 15, 2019.} Any modification of the parameters of the algorithm can only be done with the participation of a health professional. Finally, any such system must ensure the traceability of algorithmic decisions and the underlying data, and guarantee health professionals' access of those data. 

\subsubsection{United States banking laws require explanations for loan denials}

In the United States, banks are required to provide explanations for the denial of loans, whether or not the denials are based on algorithms. Regulation B requires banks to give the “principal reasons” for denying an application or taking other adverse action, including whether the bank relied on a credit score.\footnote{12 CFR Part 1002.9.} 

\subsubsection{Proposed Algorithmic Accountability Act and European Commission white paper would require a description of algorithmic design and training data}

A proposed law in the United States called the “Algorithmic Accountability Act” would require companies that deal with personal data of more than 1,000,000 consumers or have more than \$50 million in average annual gross receipts to prepare impact assessments for certain high-risk algorithms. The impact assessments would contain a "detailed description of the automated decision system, its design, its training, data, and its purpose."\footnote{Proposed Algorithmic Accountability Act, H.R. 2231, introduced April 10, 2019.} The European Commission proposes similar disclosure obligations for high-risk AI applications\citep{EuropComm2020}. 

\subsubsection{Explanation as an element of good faith and loyalty}

Legal scholars in the United States are debating whether a fiduciary duty should be imposed on large data platforms, which would include an enhanced duty to inform users about algorithmic decisions, and justify the absence of conflicts of interest \citep{balkin2015information}.  In France, a duty to provide explanations exists for contract termination \citep{houtsieff2019}, particularly for employment contracts or lease agreements. \cite{fabre92} shows that the general duty to provide information to the other contracting party is rooted in ethical considerations, and in the party’s legal duty of good faith and cooperation. 

\subsubsection{Explanations and competition law}

When a company occupies a dominant position under EU competition law, it has an enhanced duty to apply objective, transparent and non-discriminatory terms to its business partners. This can translate into a duty to make sure its terms and conditions, and business practices, are understandable and predictable to its economic partners. This constitutes a form of explanation duty. The French Competition Authority's recent €150 million sanction against Google is based in part on Google's failure to provide clear unambiguous information in its terms and conditions\footnote{French Competition Authority Decision n° 19-D-26 of December 19 2019.}. The obligations imposed on online intermediaries and search engines under the EU Platform to Business Regulation discussed above are inspired by competition law principles. But the Platform to Business Regulation extends those principles to all online intermediaries and search engines regardless of whether they occupy a dominant position under competition law. 

\section{The economics of explanations}\label{sec5}

Laws and regulations generally impose explanations when doing so is socially beneficial, that is, when the collective benefits associated with providing explanations exceed the costs. When considering algorithmic explainability, where the law has not yet determined exactly what form of explainability is required in which context, the costs and benefits of explanations need to be taken into account. The cost-benefit analysis will help determine when and how explanations should be provided, permitting various trade-offs -- such as those highlighted in the EU Platform to Business Regulation -- to be understood and managed.

\subsection{Explanations generate costs}

Like human explanations, algorithmic explanations generate costs, and should only be required in situations where the benefits of the explanations outweigh the costs \citep[p. 12]{doshi2017accountability}. According to the United Kingdom's Information Commissioner's Office, cost and resources are cited as the principal reason for not providing explanations \citep{ico2019}. The relevant question is therefore whether the benefits of a particular explanation are worth the costs. Explanations must be built to fit the particular context, and context can often be translated into a rough comparison of costs and benefits of explanations in the relevant situation. Where the stakes are high, for example investigating the cause of a serious accident, the benefits of requiring detailed explanations will be very high, because explanations could help determine who is responsible for the accident and help prevent similar accidents in the future. In a situation with lower stakes, for example finding out why Netflix recommended a particular television series, the social benefits of requiring detailed explanations will be low compared to the costs. Each situation warrants its own level of explanation, going from light explanations for algorithmic decisions with little or no consequence, to detailed explanations for high stakes algorithmic decisions. 

\medskip

\cite{doshi2017accountability} illustrate this idea using the example of a smart toaster: 
\medskip
"\textit{Requiring every AI system to explain every decision could result in less efficient systems, forced design choices, and a bias towards explainable but suboptimal outcomes. For example, the overhead of forcing a toaster to explain why it thinks the bread is ready might prevent a company from implementing a smart toaster feature -- either due to the engineering challenges or concerns about legal ramifications}" \citep[p. 12]{doshi2017accountability}.

\medskip

\cite{hleg2019high} expresses the idea in the form of a sliding scale : “\textit{the degree to which explicability is needed is highly dependent on the context and the severity of the consequences if that output is erroneous or otherwise inaccurate}.” 

\subsection{A taxonomy of explanation costs}

Many of the costs of providing explanations are difficult to quantify. Nevertheless, a critical first step to evaluating costs, even from a qualitative angle, is to break them down into categories \citep{maxwell2017smart}. Each category of cost can then be evaluated, and if necessary given a qualitative score (low, medium, high), and compared to the corresponding benefits. 

\medskip

In connection with algorithmic explanations, we have identified seven categories of costs.

\subsubsection{Design costs}

Unlike humans, algorithms do not have an implicit ability to explain their decisions after the fact. Consequently, the explanation function must be explicitly designed into the algorithm. The design costs can be high because there is no one-size-fits-all approach to explanation, and the technical approaches are varied, each approach having its advantages and drawbacks. Each kind of application may require its own approach to explainability, and the approaches may vary over different geographies based on different regulatory environments. As noted above, France has a particularly high standard for explanations relating to government algorithms, whereas the same level of explanation may not be required in other countries. Like many aspects of ethical AI, the level of algorithmic explanation will depend on context, local culture and regulation, making a global approach all but impossible. This poses a challenge to operationalizing explainable AI decision systems \citep{ico2019}.

\subsubsection{Reducing prediction accuracy}

Algorithmic accuracy and interpretability are often presented as trade-offs: increased interpretability leads to decreased accuracy and vice versa. This trade-off is largely due to the fact that most algorithms were built with only accuracy in mind. Even algorithms with explanation models built on top may lead to sacrifices in accuracy in some situations: “by requiring explanation we might cause the system to reject a solution that cannot be reduced to a human-understandable set of factors” \citep[p. 10]{doshi2017accountability}.

\medskip

In spite of technological advances, which we have examined in Section 3, most deep learning solutions do not yet incorporate explainability features, which leads some users to prefer suboptimal but easily-explained models over higher performing but more opaque models \citep[p. 12]{doshi2017accountability}.  Foregoing performance for the sake of explainability can create large social costs. For example, some estimate that criminal fund transfers account for 1-2\% of global GDP, and that less than 1\% of criminal fund transfers are currently seized through existing anti-money laundering mechanisms \citep{europol2017}. Explainability is cited as one of the reasons why banks do not introduce higher-performing AI-based system to detect criminal funds \citep{iif2018}. Even a small percentage increase in the rate of detection of hidden criminal transfers could yield enormous social benefits. In this respect, requiring fully explainable models for anti-money laundering could result in high social costs if it means that banks cannot deploy the best-performing model.

\subsubsection{Creation and storage of audit logs}

A useful source of information (and explanations) consists in stored logs relating to each algorithmic decision. Logging is one of the most straightforward ways to facilitate after-the-fact local explanations, but it is very costly \citep[pp. 651-652]{kroll2016accountable}. \cite{winfield2017case} propose that robots and autonomous systems be equipped with the equivalent of a Flight Data Recorder to continuously record sensor and relevant internal status data. The authors point out that in an autonomous vehicle, the data recorder would only be able to store the last three hours of driving data. In its white paper on AI, the European Commission does not refer explicitly to decision logs, but proposes the storage of data sets used to train and test algorithms (\citep{EuropComm2020}. The problem of storing  data is well known by telecom operators, who are obliged in some countries to store connection logs for up to two years to help law enforcement authorities investigate crimes. The costs of retaining the connection logs are significant. The costs for storing AI decision logs could potentially be much higher given the volume of data. In addition to generating new storage costs, the retention of individual decision records can also conflict with privacy legislation, which generally requires deletion of personal data as soon as possible after the relevant processing operation has been completed.  A 2006 European directive requiring telecommunication operators to retain traffic data for up to two years was found unconstitutional because it disproportionately interfered with individuals’ right to privacy.\footnote{Court of Justice of the European Union, Cases C‑293/12 and C‑594/12, Digital Rights Ireland v. Minister of Communications, April 8, 2014.} Similarly, the European Court of Justice found that imposing technical costs on a hosting provider to detect and block copyright infringing content interfered with the provider’s constitutional right to conduct a business, and therefore could only be justified if the measure passed the proportionality test.  The proportionality test requires an analysis of whether the measure is necessary to achieve the legitimate public policy objective, and in particular whether the measure is the least intrusive means available \citep{maxwell2017smart}.  

\medskip

The use of biometrics such as facial recognition illustrates the tension between explainability and the protection of personal data. In Europe, biometric access control mechanisms, e.g. fast-track border control at airports, should not generally store any personal data in the system, including information indicating why a particular person's image at the airport was not judged to be a sufficient match with the passport. While consistent with the European data protection principle of data minimization, the absence of logs can prevent the frustrated traveller from obtaining an explanation as to why his or her access was denied. The absence of logs can also make it difficult to test whether the system is biased in real-world conditions against certain types of skin color. Privacy principles can therefore conflict with the right to receive an explanation, and the right not to be subject to discriminatory AI systems. These tensions are forms of costs that need to be weighed.

\subsubsection{Violation of trade secrets}

Some of the opacity surrounding algorithmic decisions comes from deliberate corporate decisions not to share information, generally to protect trade secrets \citep{burrell2016machine}. Like other forms of property, trade secrets are constitutionally protected. The forced disclosure of source code creates costs in terms of interference with those property rights. For many situations, the source code will contribute little value to explanations. “Machine learning... is particularly ill-suited to source code analysis" \citep[p. 638]{kroll2016accountable}. 

\medskip

However, in high-stakes situations, such as the investigation of an accident, authorities will require disclosure of source code and all other relevant information that may contribute to understanding the cause of the failure. 

\medskip

The protection of trade secrets also enhances competition. A systematic disclosure of trade secrets, such as the kinds of algorithms used by data solution providers, might harm their ability to compete. Too much transparency could hurt smaller competitors more than big ones.  

\subsubsection{Conflict with security and other policy objectives}

We saw above that explanations can conflict with data protection law. Explanations can also conflict with other policy objectives, such as security. For example, the functioning of an algorithm designed to detect fraud or money laundering should remain secret in order to prevent criminals from reverse engineering the system to avoid detection. Transparency therefore must always be balanced against countervailing costs, such as costs of decreased effectiveness against fraud or cyberattacks. Search engines face the same problem: being too transparent would permit web publishers to artificially boost their own rankings by manipulating the search algorithm, which would be unfair to other web publishers.

\subsubsection{Reducing decisional flexibility in the future}

By giving an explanation, a person implicitly binds herself to apply the same rule in the future. Consequently, by requiring an explanation, we also limit the scope of future decisions by the person or entity giving the explanation \citep{schauer1995giving}. Reducing the range of options in the future can create costs because a future situation may not justify application of the same rule, and the decision-maker will be trapped by the explanation she gave in the past. 

\subsubsection{Slowing down innovation}

The United Kingdom's Information Commissioner’s Office \citep{ico2019} points out that new innovative products are being developed so quickly and frequently that legal or compliance departments are not able to provide input, including the requirement to put explainability into the product specifications. Explainability can be viewed as a compliance constraint that could increase time-to-market,  slowing  the pace of innovation.

\subsection{The benefits of explanations}

The operational benefits of explainability were mentioned in Section \ref{sec1} and \ref{sec2}. They include the ability to foster user trust in the system, and the ability to make algorithms more robust and certifiable for safety-critical applications. The societal benefits (individual rights and collective interests) were mentioned in Section \ref{sec3}. We do not propose to list each of those benefits again here. Suffice it to say that for each situation the relevant benefit from explainability will have to be defined precisely so that the benefits can be considered in relation to the costs. That the relevant benefits cannot be quantified does not prevent the benefits from being listed, and their relative importance evaluated through a scoring mechanism \citep{maxwell2017smart}. This helps put benefits in front of the corresponding costs, and facilitate a cost-benefit analysis.

\medskip

\begin{figure}[H]
\centering
\includegraphics[scale=0.7]{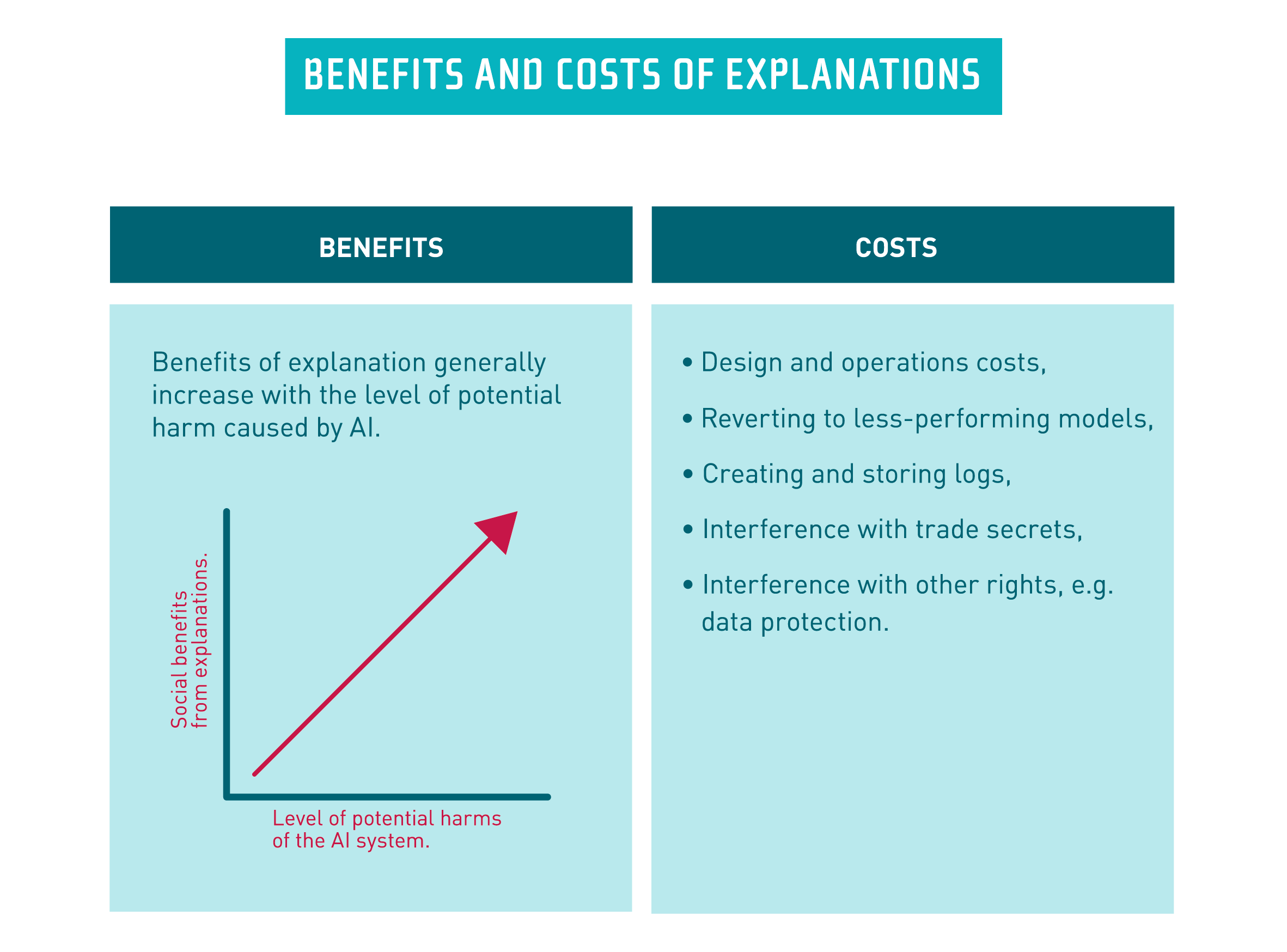}
\caption{Summary of benefits and costs of explanations}\label{VECTO_09}
\end{figure}

As a general matter, the level of potential benefits will increase with the level of potential harm that may be caused by the algorithm (see Figure \ref{VECTO_09}). An algorithm that has the potential of killing a human  (an autonomous vehicle, for example) will generate high benefits from explainability, whereas an algorithm that provides shopping recommendations will generate low benefits \citep{hleg2019high}. 

\subsection{A sliding scale of explanation solutions}

As a general rule, the higher the level of potential harm, the higher the benefits from explainability. These higher benefits will in turn justify higher explanation costs. However, even in high-stakes situations there may still be trade-offs. For example, how much individual information about an AI-powered diagnostic tool should be stored in audit logs, and how long should the logs be kept?  The storage of logs creates costs for the operator of the algorithm, but also societal costs associated with potential harms to privacy. Each case will require identifying explanation scenarios, and then assessing the benefits (and costs) of different forms of explanations in each scenario. The ideal form of explanation will be the one that maximizes net social welfare, i.e. total social benefits less total social costs.

\subsection{Four kinds of contextual factors}

%
As pointed out in the introduction, context plays a key role in determining what level of explanation may be required.  In the three-step methodology outlined in the introduction of this paper, we proposed that context be broken down into four factors:

\begin{itemize}
\item Audience/recipient factors: Who is receiving the explanation? What is their level of expertise? What are their time constraints?
\item Impact factors: What harms could the algorithm cause and how might explanations help? The higher the potential harm, the more explanation is likely to be needed.
\item Regulatory factors: What is the regulatory environment for the application? What fundamental rights are affected?
\item Operational factors: To what extent is explanation an operational imperative? For safety certification? For user trust?
\end{itemize}

These four factors will help characterize the needs for explanation, and evaluate the likely benefits. One of the most important factors is the audience: who is the explanation targeting? The human-machine interface will be a predominant consideration to how explanations are formulated. AI systems serving in a decision-support role might require different degrees of explanation than systems acting in a decision-making capacity.  For example, an expert radiologist might be able to recognize situations when an automated analysis support tool is likely to misbehave.  In such a situation, explanation might not be as necessary.

\medskip

The context of the relationship between the human user and the AI system may be as important as the relationship between two human decision-makers.  A system that acts as a partner will require different explanation than a delegate.  From a technical perspective, both of these contexts may involve decidability, but the human-machine partnership and how those decisions are to be used and interpreted plays an important role.  Similarly, accountability explanations, such as in diagnosing failure situations, will probably require a different kind of explanation than one as a part of a continuous interaction.  Moreover, the depth of explanation required may vary.  In some situations, the fact that a neural net did not recognize a given object in an image may suffice, where in others a deeper explanation of why a given object was not recognized may be required.

\medskip

This relationship will further evolve over time as a co-adaptive system~\citep{mackay90} between human user and machine.  As the user grows accustomed to the quality of the system's performance, her behavior may alter accordingly.  Ironically, as the system becomes more reliable, the user may be more likely to place undue confidence in it than a system perceived to be more unreliable (and thus in need of greater supervision).

\section{Explainability’s role in algorithmic quality control and governance}\label{sec6} 
 
Explainability has an important role in algorithmic quality control, both before the system goes to market and afterwards, because it helps bring to light weaknesses in the algorithm such as bias that would otherwise go unnoticed \citep{ico2019}.  Explainability contributes to "total product lifecycle" \citep{FDA2019} or "safety lifecycle" \citep{kurd2003} approaches to algorithmic quality and safety. 
 
\subsection{The verification and validation (V\&V) of AI-based systems}
 
The quality of machine learning models is often judged by the average accuracy rate when analyzing test data. This simple measure of quality fails to reflect weaknesses affecting the algorithm’s quality in certain situations, particularly bias and failure to generalize. The explainability solutions presented in Section \ref{sec3} can assist in identifying areas of input data where the performance of the algorithm is poor, and identify defects in the learning data that lead to bad predictions, and thereby help make the model more robust. A saliency map may reveal that a classifier is relying on irrelevant features in the training data to make predictions, for example snow in the background of photographs of wolves. Understanding areas of input data that do not belong within the training region of the deep neural network can help build so-called “safety cages” that create an alert when the input data fall outside of a range of confidence \citep{borg2019safely}. The safety cage then would redirect execution to a safe-track procedure which does not make use of the machine learning algorithm. 

\medskip
 
Traditional approaches to software verification and validation (V\&V) are poorly adapted to neural networks \citep{peterson1993foundation, borg2019safely, FDA2019}. The challenges include the non-determinism of neural network decisions, which makes it hard to demonstrate the absence of unintended functionality, and the adaptive nature of machine-learning algorithms, which makes them moving target \citep{borg2019safely,FDA2019}.  Specifying a set of requirements that comprehensively describe the behavior of a neural network is considered the most difficult challenge with regard to traditional V\&V and certification approaches \citep{borg2019safely, bhattacharyya2015certification}. The absence of complete requirements poses a problem because one of the objectives of V\&V is to compare the behavior of the software to a document that describes precisely and comprehensively the system’s intended behavior \citep{peterson1993foundation}. For neural networks, there may remain a degree of uncertainty about the output for a given input. Other barriers include the absence of detailed design documentation and the lack of interpretability of machine learning models, which challenge comprehensibility and trust which are generally required in certification processes \citep{borg2019safely}. 

\subsection{Verification of the training data}
 
Evaluation of training data is particularly important for the detection of bias \citep{lehr2017playing, peterson1993foundation}. The source and quality of the training data should be verified by human reviewers independent of the designers of the algorithm. The key characteristics of the training data, including their limitations and possible sources of bias, should be included in the global explainability user’s manual discussed above \citep{ieee2019ethically,EuropComm2020}. A full disclosure of the limitations of the training data has been imposed by at least one court in the United States as a legal explainability requirement, and that requirement is likely to become a global norm.\footnote{See the discussion of Loomis v. State of Wisconsin in Section \ref{sec5} above.} Disclosures relating to the training data would be required under the proposed Algorithmic Accountability Act in the United States, and under the European Commission's proposed approach for high-risk AI applications \citep{EuropComm2020}. Consequently the requirement of global explainability will force system developers to ensure that the quality of the training data has been evaluated and any defects corrected. For defects in training data that cannot be completely cured (e.g. bias), the defects should be prominently disclosed to users. 
 
\subsection{AI impact assessments and an independent review committee}
 
A number of approaches have emerged relating to institutional governance mechanisms for ethical AI, including the general requirement for human control \citep{EuropComm2019}. One governance approach is to require an ethical impact assessment for any AI system that is considered risky.\footnote{This approach has been proposed in the draft Algorithmic Accountability Act in the United States, the European Commission’s communication on Building Trust in Human-Centric Artificial Intelligence (2019), and the Council of Europe’s Draft Recommendation on the human rights impacts of algorithmic systems (2019).} \cite{yeung2019ai} propose an approach modeled after the ‘safety case’ framework used for regulating safety of offshore oil rigs. This approach, which is similar to the risk management framework imposed on financial institutions under the Solvency II Directive,\footnote{See Art. 44 et seq. of Directive 2009/138/EC of the European Parliament and of the Council on the taking-up and pursuit of the business of Insurance and Reinsurance (Solvency II), and     Art. 258 et seq. of Commission delegated regulation (EU) 2015/35 of 10 October 2014 supplementing such directive.} consists of developing a comprehensive risk assessment, listing all the possible things that could go wrong for the AI system -- including societal and environmental harms -- and assessing the severity of harm for each possible bad event, and the probability of each bad event occurring. The impact assessment is similar to the GDPR’s requirement for data protection impact assessments.\footnote{Regulation (EU) 2016/679, article 35.} Each harm would then be addressed by a mitigation measure to bring the risk -- considered as the product of the severity of harm multiplied by its probability -- to acceptable levels. In the case of AI systems, the risks of harm would include things like: the inability of a citizen to understand and challenge an algorithmic decision, the inability of a court or regulator to determine the cause of an accident due to the unavailability of logs, or the risk of bias from non-representative training data. Explainability solutions would be presented as mitigation measures designed to reduce each of these identified risks.
 
\medskip

Under the safety case approach, the operator of the system would bear the responsibility for developing the impact assessment and safety arguments to show that the mitigation measures are sufficient to reduce the identified risks to acceptable levels. Where residual risks remain, the user's manual accompanying the system may contain warning labels informing users about  residual risks, and advising the user not to use the algorithm in certain contexts. The operator of the system would then have to present its risk assessment to an independent body such as a safety or AI ethics regulator and convince the regulator that the risk assessment has considered all the risks, and that the mitigation measures are sufficient to address them. The regulator’s role is to stress-test the operator’s risk assessment and where appropriate improve the risk assessment and mitigation measures proposed by the operator. Once the final risk assessment and mitigation package are finalized, the measures become binding on the operator, which is similar to how risk management operates under the Solvency II Directive. The regulator then has the right to audit the operator against these requirements, and impose penalties in case of deviation. This kind of continuous oversight of safety is consistent with approaches used for safety of oil rigs, risk management in financial institutions, the regulation of AI in medical devices and for other AI systems deployed in safety-critical environments \citep{yeung2019ai,FDA2019,babic2019algorithms,kurd2003}. Instead of a risk assessment, the European Commission proposes a conformity assessment for high-risk AI applications, but the conformity assessment may well include an obligatory risk assessment by the relevant operator \citep{EuropComm2020}.

\medskip

For less risky artificial intelligence systems, an external review may not be justified. However, in that case internal committees may be set up to assess the quality of learning data and test the validity of the system based on their expertise. Just as V\&V is systematically applied in software development, a mandatory testing and validation step should be implemented in the business process, involving relevant domain experts and machine learning experts. The internal committee would have a dual responsibility: an evaluation before the system is put into production and a responsibility for handling contentious cases that are reported over time by the field. 

\section{Conclusion: Context-specific AI explanations by design}\label{concl}

Explainability is both an operational and ethical requirement. The operational needs for explainability are driven by the need to increase robustness, particularly for safety-critical applications, as well as enhance acceptance by system users. The ethical needs for explainability address harms to fundamental rights and other societal interests which may be insufficiently addressed by the purely operational requirements.

\medskip

Research on explainable AI focuses first and foremost on operational requirements: how to make systems more robust and safe through explainability?  The use of hybrid solutions, combining machine learning and symbolic AI, is a promising field of research for safety-critical applications, and applications such as medicine where important bodies of domain knowledge must be associated with algorithmic decisions. Explainability will increasingly become a performance criterion for algorithms, and included by design. 

\medskip

Regulation of AI explainability remains largely unexplored territory, the most ambitious efforts to date being the French law on the explainability of government algorithms and the EU regulation on Platform to Business relations. However, even in those instances, the law leaves many aspects of explainability open to interpretation. The form of explanation and the level of detail will be driven by the four categories of contextual factors described in this paper: audience/recipient factors, impact factors, regulatory factors, and operational factors. The level of detail of explanations -- global or local -- would follow a sliding scale depending on the context, and the costs and benefits at stake. One of the biggest costs of local explanations will relate to storage of individual decision logs. The kind of information stored in the  logs, and the duration of storage, will be key questions to address when determining the right level of explainability. AI impact assessments, which are likely to be imposed by law for risky AI systems, are the right place to weigh these contextual factors in order to design right-fitting explainability solutions.

\medskip

As technical solutions to explainability converge toward hybrid AI approaches, we can expect that the trade-off between explainability and performance will become less acute. Explainability will become part of performance. Also, as explainability becomes a requirement for safety certification, we can expect alignment between operational/safety needs for explainability and ethical/human rights needs for explainability. Some of the solutions for operational explainability may serve both purposes.  

\bibliographystyle{apalike} 
\bibliography{refs}

\begin{thebibliography}{}

\bibitem[Al-Shedivat et~al., 2017]{cen}
Al-Shedivat, M., Dubey, A., and Xing, E.~P. (2017).
\newblock Contextual explanation networks.
\newblock {\em arXiv preprint arXiv:1705.10301}.

\bibitem[Alvarez-Melis and Jaakkola, 2017]{socrat}
Alvarez-Melis, D. and Jaakkola, T.~S. (2017).
\newblock A causal framework for explaining the predictions of black-box
  sequence-to-sequence models.
\newblock {\em arXiv preprint arXiv:1707.01943}.

\bibitem[Babic et~al., 2019]{babic2019algorithms}
Babic, B., Gerke, S., Evgeniou, T., and Cohen, I.~G. (2019).
\newblock Algorithms on regulatory lockdown in medicine.
\newblock {\em Science}, 366(6470):1202--1204.

\bibitem[Balkin, 2015]{balkin2015information}
Balkin, J.~M. (2015).
\newblock Information fiduciaries and the first amendment.
\newblock {\em UCDL Rev.}, 49:1183.

\bibitem[Bang et~al., 2019]{vibi}
Bang, S., Xie, P., Wu, W., and Xing, E. (2019).
\newblock Explaining a black-box using deep variational information bottleneck
  approach.
\newblock {\em arXiv preprint arXiv:1902.06918}.

\bibitem[Bhattacharyya et~al., 2015]{bhattacharyya2015certification}
Bhattacharyya, S., Cofer, D., Musliner, D., Mueller, J., and Engstrom, E.
  (2015).
\newblock Certification considerations for adaptive systems.
\newblock In {\em 2015 International Conference on Unmanned Aircraft Systems
  (ICUAS)}, pages 270--279. IEEE.

\bibitem[Borg et~al., 2019]{borg2019safely}
Borg, M., Englund, C., Wnuk, K., Duran, B., Levandowski, C., Gao, S., Tan, Y.,
  Kaijser, H., L{\"o}nn, H., and T{\"o}rnqvist, J. (2019).
\newblock Safely entering the deep: A review of verification and validation for
  machine learning and a challenge elicitation in the automotive industry.
\newblock {\em Journal of Automotive Software Engineering}, 1(1):1--19.

\bibitem[Breiman, 1984]{breiman1984algorithm}
Breiman, L. (1984).
\newblock Algorithm cart.
\newblock {\em Classification and Regression Trees. California Wadsworth
  International Group, Belmont, California}.

\bibitem[Breiman, 2001]{Breiman2001}
Breiman, L. (2001).
\newblock Random forests.
\newblock {\em Machine Learning}, 45(1):5--32.

\bibitem[Burrell, 2016]{burrell2016machine}
Burrell, J. (2016).
\newblock How the machine ‘thinks’: Understanding opacity in machine
  learning algorithms.
\newblock {\em Big Data \& Society}, 3(1):2053951715622512.

\bibitem[Chen et~al., 2019]{chen2019looks}
Chen, C., Li, O., Tao, D., Barnett, A., Rudin, C., and Su, J.~K. (2019).
\newblock This looks like that: deep learning for interpretable image
  recognition.
\newblock In {\em Advances in Neural Information Processing Systems}, pages
  8928--8939.

\bibitem[Clancey, 1983]{clancey1983epistemology}
Clancey, W.~J. (1983).
\newblock The epistemology of a rule-based expert system—a framework for
  explanation.
\newblock {\em Artificial intelligence}, 20(3):215--251.

\bibitem[Coglianese and Lehr, 2018]{coglianese2018transparency}
Coglianese, C. and Lehr, D. (2018).
\newblock Transparency and algorithmic governance.
\newblock {\em Administrative Law Review}, 71:1.

\bibitem[Cohen et~al., 2017]{cohen2017tensorlog}
Cohen, W.~W., Yang, F., and Mazaitis, K.~R. (2017).
\newblock Tensorlog: Deep learning meets probabilistic dbs.

\bibitem[Cortes and Vapnik, 1995]{cortes1995support}
Cortes, C. and Vapnik, V. (1995).
\newblock Support-vector networks.
\newblock {\em Machine learning}, 20(3):273--297.

\bibitem[Dessalles, 2019]{Dessalles2019}
Dessalles, J.-L. (2019).
\newblock {\em Des intelligences TRÈS artificielles}.
\newblock Odile Jacob.

\bibitem[Doshi-Velez et~al., 2017]{doshi2017accountability}
Doshi-Velez, F., Kortz, M., Budish, R., Bavitz, C., Gershman, S., O'Brien, D.,
  Schieber, S., Waldo, J., Weinberger, D., and Wood, A. (2017).
\newblock Accountability of ai under the law: The role of explanation.
\newblock {\em arXiv preprint arXiv:1711.01134}.

\bibitem[d’Alch{\'e} Buc et~al., 1994]{d1994rule}
d’Alch{\'e} Buc, F., Andr{\'e}s, V., and Nadal, J.-P. (1994).
\newblock Rule extraction with fuzzy neural network.
\newblock {\em International journal of neural systems}, 5(01):1--11.

\bibitem[Ernest et~al., 2016]{ernest2016genetic}
Ernest, N., Carroll, D., Schumacher, C., Clark, M., Cohen, K., and Lee, G.
  (2016).
\newblock Genetic fuzzy based artificial intelligence for unmanned combat
  aerial vehicle control in simulated air combat missions.
\newblock {\em Journal of Defense Management}, 6(1):2167--0374.

\bibitem[{European Commission}, 2019]{EuropComm2019}
{European Commission} (2019).
\newblock Communication from the commission to the european parliament, the
  council, the european economic and social committee and the committee of the
  regions - building trust in human centric artificial intelligence
  (com(2019)168).
\newblock Technical report.

\bibitem[{European Commission}, 2020]{EuropComm2020}
{European Commission} (2020).
\newblock White paper on artificial intelligence - a european approach to
  excellence and trust (com(2020)65 final).
\newblock Technical report.

\bibitem[Europol, 2017]{europol2017}
Europol (2017).
\newblock From suspicion to action - converting financial intelligence into
  greater operational impact.
\newblock Technical report, Europol.

\bibitem[Fabre-Magnan, 1992]{fabre92}
Fabre-Magnan, M. (1992).
\newblock De l’obligation d’information dans les contrats - essai d’une
  théorie.
\newblock {\em Revue de droit d’Assas}.

\bibitem[Felici et~al., 2013]{felici2013accountability}
Felici, M., Koulouris, T., and Pearson, S. (2013).
\newblock Accountability for data governance in cloud ecosystems.
\newblock In {\em 2013 IEEE 5th International Conference on Cloud Computing
  Technology and Science}, volume~2, pages 327--332. IEEE.

\bibitem[Fisher, 1936]{fisher1936linear}
Fisher, R. (1936).
\newblock Linear discriminant analysis.
\newblock {\em Ann. Eugenics}, 7:179.

\bibitem[Guidotti et~al., 2018]{guidotti2018survey}
Guidotti, R., Monreale, A., Ruggieri, S., Turini, F., Giannotti, F., and
  Pedreschi, D. (2018).
\newblock A survey of methods for explaining black box models.
\newblock {\em ACM computing surveys (CSUR)}, 51(5):93.

\bibitem[Haugeland, 1985]{Haugeland1985}
Haugeland, J. (1985).
\newblock {\em Artificial Intelligence: The Very Idea}.
\newblock Massachusetts Institute of Technology, USA.

\bibitem[HLEG, 2019]{hleg2019high}
HLEG, A. (2019).
\newblock High-level expert group on artificial intelligence.
\newblock {\em Ethics Guidelines for Trustworthy AI}.

\bibitem[Houtsieff, 2019]{houtsieff2019}
Houtsieff, D. (2019).
\newblock La motivation en droit des contrats.
\newblock {\em Revue de droit d’Assas}, 19.

\bibitem[ICO, 2019]{ico2019}
ICO (2019).
\newblock Project explain interim report.
\newblock Technical report, Information Commissioner’s Office.

\bibitem[IEEE, 2019]{ieee2019ethically}
IEEE (2019).
\newblock Ethically aligned design: A vision for prioritizing human well-being
  with autonomous and intelligent systems.
\newblock {\em IEEE Global Initiative on Ethics of Autonomous and Intelligent
  Systems}.

\bibitem[{Institute of International Finance}, 2018]{iif2018}
{Institute of International Finance} (2018).
\newblock Machine learning in anti-money laundering – summary report.
\newblock Technical report.

\bibitem[Kim et~al., 2017]{kim17tcav}
Kim, B., Wattenberg, M., Gilmer, J., Cai, C., Wexler, J., Viegas, F., and
  Sayres, R. (2017).
\newblock Interpretability beyond feature attribution: Quantitative testing
  with concept activation vectors (tcav).
\newblock {\em arXiv preprint arXiv:1711.11279}.

\bibitem[Kindermans et~al., 2017]{kindermans2017unreliability}
Kindermans, P.-J., Hooker, S., Adebayo, J., Alber, M., Schütt, K.~T., Dähne,
  S., Erhan, D., and Kim, B. (2017).
\newblock The (un)reliability of saliency methods.

\bibitem[Kroll et~al., 2016]{kroll2016accountable}
Kroll, J.~A., Barocas, S., Felten, E.~W., Reidenberg, J.~R., Robinson, D.~G.,
  and Yu, H. (2016).
\newblock Accountable algorithms.
\newblock {\em U. Pa. L. Rev.}, 165:633.

\bibitem[Kuhn and Kacker, 2019]{kuhn2019application}
Kuhn, R. and Kacker, R. (2019).
\newblock An application of combinatorial methods for explainability in
  artificial intelligence and machine learning (draft).
\newblock Technical report, National Institute of Standards and Technology.

\bibitem[Kurd and Kelly, 2003]{kurd2003}
Kurd, Z. and Kelly, T. (2003).
\newblock Safety lifecycle for developing safety critical artificial neural
  networks.
\newblock In Anderson, S., Felici, M., and Littlewood, B., editors, {\em
  Computer Safety, Reliability, and Security}, pages 77--91, Berlin,
  Heidelberg. Springer Berlin Heidelberg.

\bibitem[Lee et~al., 2019a]{roll}
Lee, G.-H., Alvarez-Melis, D., and Jaakkola, T.~S. (2019a).
\newblock Towards robust, locally linear deep networks.
\newblock In {\em International Conference on Learning Representations}.

\bibitem[Lee et~al., 2019b]{game}
Lee, G.-H., Jin, W., Alvarez-Melis, D., and Jaakkola, T.~S. (2019b).
\newblock Functional transparency for structured data: a game-theoretic
  approach.
\newblock {\em arXiv preprint arXiv:1902.09737}.

\bibitem[Lehr and Ohm, 2017]{lehr2017playing}
Lehr, D. and Ohm, P. (2017).
\newblock Playing with the data: what legal scholars should learn about machine
  learning.
\newblock {\em UCDL Rev.}, 51:653.

\bibitem[Lundberg and Lee, 2017]{shap}
Lundberg, S.~M. and Lee, S.-I. (2017).
\newblock A unified approach to interpreting model predictions.
\newblock In {\em Advances in Neural Information Processing Systems}, pages
  4765--4774.

\bibitem[Mackay, 1990]{mackay90}
Mackay, W.~E. (1990).
\newblock {\em Users and Customizable Software: A Co-Adaptive Phenomenon}.
\newblock PhD thesis, Massechusetts Institute of Technology.

\bibitem[Mascarilla, 1997]{Mascarilla1997}
Mascarilla, L. (1997).
\newblock {Fuzzy Rules Extraction and Redunduncy Elimination: an Application to
  Remote Sensing Image Analysis}.
\newblock {\em International Journal of Intelligent Systems},
  12(11/12):793--818.

\bibitem[Maxwell, 2017]{maxwell2017smart}
Maxwell, W.~J. (2017).
\newblock {\em Smart (er) Internet Regulation Through Cost-Benefit Analysis:
  Measuring harms to privacy, freedom of expression, and the internet
  ecosystem}.
\newblock Presses des Mines via OpenEdition.

\bibitem[Melis and Jaakkola, 2018]{senn}
Melis, D.~A. and Jaakkola, T. (2018).
\newblock Towards robust interpretability with self-explaining neural networks.
\newblock In {\em Advances in Neural Information Processing Systems}, pages
  7775--7784.

\bibitem[Minsky and Papert, 1969]{minsky1969}
Minsky, M. and Papert, S. (1969).
\newblock {\em Perceptrons}.
\newblock M.I.T. Press.

\bibitem[OECD, 2019a]{OECD2019a}
OECD (2019a).
\newblock {\em Artificial Intelligence in Society}.

\bibitem[OECD, 2019b]{OECD2019b}
OECD (2019b).
\newblock {\em Recommendation of the Council on Artificial Intelligence}.

\bibitem[Peterson, 1993]{peterson1993foundation}
Peterson, G.~E. (1993).
\newblock Foundation for neural network verification and validation.
\newblock In {\em Science of Artificial Neural Networks II}, volume 1966, pages
  196--207. International Society for Optics and Photonics.

\bibitem[Ribeiro et~al., 2016a]{lime}
Ribeiro, M.~T., Singh, S., and Guestrin, C. (2016a).
\newblock Why should i trust you?: Explaining the predictions of any
  classifier.
\newblock In {\em Proceedings of the 22nd ACM SIGKDD international conference
  on knowledge discovery and data mining}, pages 1135--1144. ACM.

\bibitem[Ribeiro et~al., 2016b]{ribeiro2016}
Ribeiro, M.~T., Singh, S., and Guestrin, C. (2016b).
\newblock "why should i trust you?": Explaining the predictions of any
  classifier.

\bibitem[Rosenblatt, 1957]{rosenblatt1957perceptron}
Rosenblatt, F. (1957).
\newblock {\em The perceptron, a perceiving and recognizing automaton Project
  Para Report No. 85-460-1}.
\newblock Cornell Aeronautical Laboratory.

\bibitem[Rumelhart et~al., 1986]{rumelhart1986learning}
Rumelhart, D.~E., Hinton, G.~E., and Williams, R.~J. (1986).
\newblock Learning representations by back-propagating errors.
\newblock {\em nature}, 323(6088):533--536.

\bibitem[Schapire, 1990]{schapire1990strength}
Schapire, R.~E. (1990).
\newblock The strength of weak learnability.
\newblock {\em Machine learning}, 5(2):197--227.

\bibitem[Schauer, 1995]{schauer1995giving}
Schauer, F. (1995).
\newblock Giving reasons.
\newblock {\em Stanford Law Review}, pages 633--659.

\bibitem[Selbst and Barocas, 2018]{Selbst2018}
Selbst, A. and Barocas, S. (2018).
\newblock The intuitive appeal of explainable machines.
\newblock {\em SSRN Electronic Journal}, 87.

\bibitem[Selsam et~al., 2018]{selsam2018learning}
Selsam, D., Lamm, M., Bünz, B., Liang, P., de~Moura, L., and Dill, D.~L.
  (2018).
\newblock Learning a sat solver from single-bit supervision.

\bibitem[Selvaraju et~al., 2017]{gradcam}
Selvaraju, R.~R., Cogswell, M., Das, A., Vedantam, R., Parikh, D., and Batra,
  D. (2017).
\newblock Grad-cam: Visual explanations from deep networks via gradient-based
  localization.
\newblock In {\em Proceedings of the IEEE International Conference on Computer
  Vision}, pages 618--626.

\bibitem[Shortliffe and Buchanan, 1975]{Shortliffe1975}
Shortliffe, E.~H. and Buchanan, B.~G. (1975).
\newblock A model of inexact reasoning in medicine.
\newblock {\em Mathematical Biosciences}, 23(3):351 -- 379.

\bibitem[Simonyan et~al., 2013]{saliency}
Simonyan, K., Vedaldi, A., and Zisserman, A. (2013).
\newblock Deep inside convolutional networks: Visualising image classification
  models and saliency maps.
\newblock {\em arXiv preprint arXiv:1312.6034}.

\bibitem[Smilkov et~al., 2017]{smoothgrad}
Smilkov, D., Thorat, N., Kim, B., Vi{\'e}gas, F., and Wattenberg, M. (2017).
\newblock Smoothgrad: removing noise by adding noise.
\newblock {\em arXiv preprint arXiv:1706.03825}.

\bibitem[Springenberg et~al., 2014]{springenberg2014striving}
Springenberg, J.~T., Dosovitskiy, A., Brox, T., and Riedmiller, M. (2014).
\newblock Striving for simplicity: The all convolutional net.
\newblock {\em arXiv preprint arXiv:1412.6806}.

\bibitem[Stone, 1977]{stone1977consistent}
Stone, C.~J. (1977).
\newblock Consistent nonparametric regression.
\newblock {\em The Annals of Statistics}, pages 595--620.

\bibitem[Sundararajan et~al., 2017]{ig}
Sundararajan, M., Taly, A., and Yan, Q. (2017).
\newblock Axiomatic attribution for deep networks.
\newblock In {\em Proceedings of the 34th International Conference on Machine
  Learning-Volume 70}, pages 3319--3328. JMLR. org.

\bibitem[Swartout et~al., 1991]{swartout1991explanations}
Swartout, W., Paris, C., and Moore, J. (1991).
\newblock Explanations in knowledge systems: Design for explainable expert
  systems.
\newblock {\em IEEE Expert}, 6(3):58--64.

\bibitem[Thomas et~al., 2019]{Thomas2019}
Thomas, P.~S., Castro~da Silva, B., Barto, A.~G., Giguere, S., Brun, Y., and
  Brunskill, E. (2019).
\newblock Preventing undesirable behavior of intelligent machines.
\newblock {\em Science}, 366(6468):999--1004.

\bibitem[Tishby et~al., 2000]{tishby2000information}
Tishby, N., Pereira, F.~C., and Bialek, W. (2000).
\newblock The information bottleneck method.
\newblock {\em arXiv preprint physics/0004057}.

\bibitem[Towell and Shavlik, 1994]{towell1994knowledge}
Towell, G.~G. and Shavlik, J.~W. (1994).
\newblock Knowledge-based artificial neural networks.
\newblock {\em Artificial intelligence}, 70(1-2):119--165.

\bibitem[{US Food and Drug Administration}, 2019]{FDA2019}
{US Food and Drug Administration} (2019).
\newblock Proposed regulatory framework for modifications to artificial
  intelligence/machine learning (ai/ml)-based software as a medical device.
\newblock Technical report.

\bibitem[Wachter et~al., 2017]{wachter2017counterfactual}
Wachter, S., Mittelstadt, B., and Russell, C. (2017).
\newblock Counterfactual explanations without opening the black box: Automated
  decisions and the gpdr.
\newblock {\em Harv. JL \& Tech.}, 31:841.

\bibitem[Waltl and Vogl, 2018]{waltl2018explainable}
Waltl, B. and Vogl, R. (2018).
\newblock Explainable artificial intelligence—the new frontier in legal
  informatics.
\newblock {\em Jusletter IT}, 4:1--10.

\bibitem[Welling, 2015]{welling2015ml}
Welling, M. (2015).
\newblock Are ml and statistics complementary?
\newblock In {\em IMS-ISBA Meeting on ‘Data Science in the Next 50 Years}.

\bibitem[Winfield and Jirotka, 2017]{winfield2017case}
Winfield, A.~F. and Jirotka, M. (2017).
\newblock The case for an ethical black box.
\newblock In {\em Annual Conference Towards Autonomous Robotic Systems}, pages
  262--273. Springer.

\bibitem[Wortham et~al., 2016]{wortham2016does}
Wortham, R.~H., Theodorou, A., and Bryson, J.~J. (2016).
\newblock What does the robot think? transparency as a fundamental design
  requirement for intelligent systems.
\newblock In {\em Ijcai-2016 ethics for artificial intelligence workshop}.

\bibitem[Yeung et~al., 2019]{yeung2019ai}
Yeung, K., Howes, A., and Pogrebna, G. (2019).
\newblock Ai governance by human rights-centred design, deliberation and
  oversight: An end to ethics washing.
\newblock {\em The Oxford Handbook of AI Ethics, Oxford University Press
  (2019)}.

\bibitem[Zhang et~al., 2018]{icnn}
Zhang, Q., Nian~Wu, Y., and Zhu, S.-C. (2018).
\newblock Interpretable convolutional neural networks.
\newblock In {\em Proceedings of the IEEE Conference on Computer Vision and
  Pattern Recognition}, pages 8827--8836.

\end{thebibliography}

\pagebreak

\appendix
\section{Appendix: Glossary of AI explainability terms}\label{appendixA}

\begin{center}
\begin{tabular}{p{0.2\textwidth}p{0.8\textwidth}}\hline
Concept & Definition \\ \hline

-ability & “ability, inclination, or suitability for a specified action or condition” (American Heritage Dictionary, Fifth Edition) \\ \hline

Accountability & "Accountability focuses on being able to place the onus on the appropriate organisations or individuals for the proper functioning of AI systems. Criteria for accountability include respect for principles of human values and fairness, transparency, robustness and safety." \citep[p. 99]{OECD2019a}

\smallskip

"Liability to account for and answer for one’s conduct; judgment of blameworthiness; obligation to provide a satisfactory answer to an external oversight agent." \citep{ieee2019ethically}

\smallskip

"A set of mechanisms, practices and attributes that sum to a governance structure which “consists of accepting responsibility for the stewardship of personal and/or confidential data with which it [data organization] is entrusted in a cloud environment, for processing, storing, sharing, deleting and otherwise using data according to contractual and legal requirements from the time it is collect- ed until when the data are destroyed (including on- ward transfer to and from third parties). Accountability involves committing to legal and ethical obligations, policies, procedures and mechanism, explaining and demonstrating ethical implementation to internal and external stakeholders and remedying any failure to act properly" \citep{felici2013accountability}. 

\smallskip

“Accountability - the need to have corporate governance measures in place to appropriately manage the whole AI decision process” \citep[p. 31]{ico2019}.

\smallskip

Accountable: “expected or required to account for one’s actions; answerable. Capable of being explained” (American Heritage Dictionary, Fifth Edition) \\ \hline

Accuracy & “Accuracy: to which extent the model accurately predict unseen instances.” \citep[p. 7]{guidotti2018survey} \\ \hline

Auditability & Audit: “to examine or evaluate (something) closely” (American Heritage Dictionary, Fifth Edition) \\ \hline

Comprehensiblity & “According to the literature, we refer to interpretability also with the name comprehensibility.” \citep[p. 7]{guidotti2018survey} \\ \hline

Domain theory & “In machine learning, a domain theory is a collection of rules that describes task-specific inferences that can be drawn from the given facts. For classification problems, a domain theory can be used to prove whether or not an object is a member of a particular class.” \citep[p. 119]{towell1994knowledge} \\ \hline

\end{tabular}
\end{center}

\pagebreak

\begin{center}
\begin{tabular}{p{0.2\textwidth}p{0.8\textwidth}}\hline
Concept & Definition \\ \hline

Empirical learning & “Empirical learning systems inductively generalize specific examples. Thus, they require little theoretical knowledge about the problem domain; instead they require a large library of examples.” \citep[pp. 121-122]{towell1994knowledge} \\ \hline
 
 \smallskip

Explainability and explanations & “In this paper, when we use the term explanation, we shall mean a human-interpretable description of the process by which a decision-maker took a particular set of inputs and reached a particular conclusion. In addition to this formal definition of an explanation, and explanation must also have the correct type of content in order for it to be useful. As a governing principle for the content an explanation should contain, we offer the following: an explanation should permit an observer to determine the extent to which a particular input was determinative or influential on the output. Another way of formulating this principle is to say that an explanation should be able to answer at least one of the following questions:
What were the main factors in a decision?
Would changing a certain factor have changed the decision?....
Why did two similar-looking cases get different decisions, or vice versa?”
\citep[p. 3]{doshi2017accountability}

\smallskip

“... explanation is distinct from transparency. Explanation does not require knowing the flow of bits through an AI system, no more than explanation from humans requires knowing the flow of signals through neurons (neither of which would be interpretable to a human!). Instead, explanation, as required under the law, as outlined in Section 2, is about answering how certain factors were used to come to the outcome in a specific situation.” \citep[pp. 6-7]{doshi2017accountability}

\smallskip

“We argue that regulation around explanation from AI systems should consider the explanation system as distinct from the AI system….the designer of the explanation system must output a human-interpretable rule ex() that takes in the same input $x$ and outputs a prediction $~y$. To be locally faithful under counterfactual reasoning formally means that the predictions $~y$ and $^y$ are the same under small perturbations of the input $x$.” \citep[p. 7]{doshi2017accountability}

\smallskip

Explain: “to make plain or comprehensible” (American Heritage Dictionary, Fifth Edition)

\end{tabular}
\end{center}

\pagebreak

\begin{center}
\begin{tabular}{p{0.2\textwidth}p{0.8\textwidth}}\hline
Concept & Definition \\ \hline

Explainability and explanations & Explainability concerns the ability to explain both the technical processes of an AI system and the related human decisions (e.g. application areas of a system). Technical explainability requires that the decisions made by an AI system can be understood and traced by human beings. Moreover, trade-offs might have to be made between enhancing a system's explainability (which may reduce its accuracy) or increasing its accuracy (at the cost of explainability). Whenever an AI system has a significant impact on people’s lives, it should be possible to demand a suitable explanation of the AI system’s decision-making process. Such explanation should be timely and adapted to the expertise of the stakeholder concerned (e.g. layperson, regulator or researcher). In addition, explanations of the degree to which an AI system influences and shapes the organisational decision-making process, design choices of the system, and the rationale for deploying it, should be available (hence ensuring business model transparency). \citep[p. 18]{hleg2019high} \\ \hline

Explanation -- black box explanation & “Given a black box model solving a classification problem, the black box explanation problem consists in providing an interpretable and transparent model which is able to mimic the behavior of the black box and which is also understandable by humans (see Figure 4). In other words, the interpretable model approximating the black box must be globally interpretable.” \citep[p. 13]{guidotti2018survey} \\ \hline

Feature extraction & “There are also ways of simplifying machine learning models such as ‘feature extraction’ an approach that analyses what features actually matter to the classification outcome removing all other features from the model.” \citep{burrell2016machine} \\ \hline

Fidelity & “The fidelity captures how much is good an interpretable model in the mimic of the behavior of a black-box.” \citep[p. 7]{guidotti2018survey} \\ \hline

Fidelity -- local fidelity & “...for an explanation to be meaningful it must at least be locally faithful, i.e. it must correspond to how the model behaves in the vicinity of the instance being predicted.” \citep{ribeiro2016}

\\ \hline

Fidelity – global fidelity & “We note that local fidelity does not imply global fidelity: features that are globally important may not be important in the local context, and vice versa. While global fidelity would imply local fidelity, identifying globally faithful explanations that are interpretable remains a challenge for complex models.” \citep{ribeiro2016}  \\ \hline

\end{tabular}
\end{center}

\pagebreak

\begin{center}
\begin{tabular}{p{0.2\textwidth}p{0.8\textwidth}}\hline
Concept & Definition \\ \hline

Inscrutability & “... the property of inscrutability suggests that models available for direct inspection may defy understanding...” \citep[p. 1091]{Selbst2018}

\smallskip
 
“We define this difficulty as ‘inscrutability’ – a situation in which the rules that govern decision-making are so complex, numerous, and interdependent that they defy practical inspection and resist comprehension” \citep[p. 1094]{Selbst2018}
 \\ \hline

Interpretability & “On the machine learning side, the subfield of “interpretability” – within which researchers have been attempting to find ways to understand complex models – is over thirty years old.” \citep[p. 1099]{Selbst2018} 

\smallskip

“...in data mining and machine learning, interpretability is defined as the ability to explain or to provide the meaning in understandable terms to a human.” \citep[p. 5]{guidotti2018survey} 

\smallskip

“An essential criterion for explanations is that they must be interpretable, i.e., provide qualitative understanding between the input variables and the response. We note that interpretability must take into account the user’s limitations.” \citep[p. 2]{ribeiro2016}

\smallskip

“interpretability summarizes every effort that is made by humans or machines to provide descriptive information, i.e. visualizations, local explanations, that enable humans to understand the decisions made by ADM.” \citep{waltl2018explainable}

\smallskip

Interpret: “to explain the meaning of; to translate from one language to another” (American Heritage Dictionary, Fifth Edition) \\ \hline

Global interpretability (or explainability) & “... the property of inscrutability suggests that models available for direct inspection may defy understanding...” \citep[p. 1091]{Selbst2018} \\ \hline

Local interpretability (or explainability) & “... the property of inscrutability suggests that models available for direct inspection may defy understanding...” \citep[p. 1091]{Selbst2018} \\ \hline

Hand-built classifiers & “Hand-built classifiers correspond to teaching by giving a person a domain theory without an extensive set of examples; one could call this learning by being told.” \citep[p. 120]{towell1994knowledge} \\ \hline

Hybrid systems & “... a “hybrid” system that effectively combines a hand-built classifier with an empirical learning algorithm might be like a student who is taught using a combination of theoretical information and examples.” \citep[p. 120]{towell1994knowledge} \\ \hline

\end{tabular}
\end{center}

\pagebreak

\begin{center}
\begin{tabular}{p{0.2\textwidth}p{0.8\textwidth}}\hline
Concept & Definition \\ \hline

KBANN (Knowledge-Based Artificial Neural Networks) & “KBANN (Knowledge-Based Artificial Neural Networks) is a hybrid learning system built on top of connectionist learning techniques. It maps problem-specific “domain theories”, represented in propositional logic, into neural networks and then refines this reformulated knowledge using backpropagation.” \citep[p. 119]{towell1994knowledge}  \\ \hline

Logging & “One extremely straightforward and very commonly used form of dynamic program review comes from the practice of logging, or recording certain program actions in a file either immediately before or immediately after they have taken place. Analysis of log messages is among the easiest and is perhaps the most common type of functional review performed on most software programs. However, analyzing program logs requires that programs be written to log when they perform events which might be interesting for analysis (and that they log enough information about those events to actually perform the analysis in question)....Because of this, audit logs meant to record sensitive actions requiring reliable review are generally access controlled or sent to special restricted remote systems dedicated to receiving logging data.” \citep[pp. 651-652]{kroll2016accountable}

\smallskip

Log: “a record, as of the performance of a machine or the progress of an undertaking” (American Heritage Dictionary, Fifth Edition). \\ \hline

Monotonicity & “A predictor respecting the monotonicity principle is, for example, a predictor where the increase of the values of a numerical attribute tends to either increase or decrease in a monotonic way the probability of a record of being member of a class.” \citep[p. 7]{guidotti2018survey}\\ \hline

Non-intuitiveness & “...non intuitiveness suggests that even where models are understandable, they may rest on apparent statistical relationships that defy intuition.” \citep[pp. 1091]{Selbst2018}

“The problem in such cases is not inscrutability, but an inability to weave a sensible story to account for the statistical relationships in the model. Although the statistical relationship that serves as the basis for decision-making might be readily identifiable, that relationship pay defy intuitive expectations about the relevance of certain criteria to the decision.” \citep[pp. 1096-1097]{Selbst2018} \\ \hline

\end{tabular}
\end{center}

\pagebreak

\begin{center}
\begin{tabular}{p{0.2\textwidth}p{0.8\textwidth}}\hline
Concept & Definition \\ \hline

Opacity & “Three distinct forms of opacity include: (1) opacity as intentional corporate or institutional self-protection and concealment and, along with it, the possibility of knowing deception; (2) opacity stemming from the current state of affairs where writing (and reading) code is a specialist skill and; (3) an opacity that stems from the mismatch between mathematical optimization in high-dimensionality characteristic of machine learning and the demands of human-scale reasoning and styles of semantic interpretation.” \citep{burrell2016machine}\\ \hline

Partial dependence plots (also variable importance plots) & “More specifically, they produce plots -- often called partial dependence or individual conditional expectation plots -- that graph the outcome variable as a function of a given input variable.” \citep[p. 710]{lehr2017playing}

\smallskip

“The output of such methods is known as a variable importance plot, which displays graphically the relative importances of different input variables.” \citep[p. 708]{lehr2017playing}

\smallskip

“These plots help in visualizing and understanding the relationship between the outcome of a black box and the input in a reduced feature space.” \citep[p. 17]{guidotti2018survey}\\ \hline

Procedural regularity & “We show that computer systems can be designed to prove to oversight authorities and the public that decisions were made under an announced set of rules consistently applied in each case, a condition we call procedural regularity... 

\smallskip

Procedural regularity ensures that a decision was made using consistently applied standards and practices. It does not, however, guarantee that such practices are themselves good policy.” \citep[pp. 637-638]{kroll2016accountable}\\ \hline

Robust & “The system continues to provide useful results even if some of the inputs are missing or estimated.” \citep[p. 204]{peterson1993foundation}

“In order to test for robustness, the data can be modified and the response checked. For example, we can find out what the network does if it is given an outlier; or we can try running the network when some of the data is missing, or if the user estimates the value of the missing data.” \citep[p. 204]{peterson1993foundation}\\ \hline

Saliency masks & ”An efficient way of pointing out what causes a certain outcome, especially when images or texts are treated, consists in using “masks” visually highlighting the determining aspects of the record analyzed. They are generally used to explain deep neural networks.” \citep[p. 17]{guidotti2018survey}\\ \hline

\end{tabular}
\end{center}

\pagebreak

\begin{center}
\begin{tabular}{p{0.2\textwidth}p{0.8\textwidth}}\hline
Concept & Definition \\ \hline

Simulatability & “As computer scientist Zachary Lipton explains, simulatability -- the ability to practically execute a model in one’s mind -- is an important form of understanding a model.” \citep[p. 1096]{Selbst2018}

\smallskip

Simulation is a remarkably flat and functional definition of understanding, but it seems like a minimum requirement for any more elaborate definition. This notion of understanding has nothing to say about why the model behaves the way it does; it is simply a way to account for the facility with which a person can play out how a model would behave under different circumstances. When models are too complex for humans to perform this task, they have reached the point of inscrutability.” \citep[p. 1096]{Selbst2018} \\ \hline

\smallskip

Source-code analysis & “Perhaps the most obvious approach is to disclose a system’s source code, but this is at best a partial solution to the problem of accountability for automated decisions… Machine learning, one increasingly popular approach to automated decision-making, is particularly ill-suited to source code analysis because it involves situations where the decisional rule itself emerges automatically from the specific data under analysis, sometimes in ways that no human can explain.” \citep[p. 638]{kroll2016accountable} \\ \hline

Testing -- black box testing, white box testing & “Dynamic testing can be divided into ‘black-box testing,’ which considers only the inputs and outputs of a system or component, and ‘white-box testing,’ in which the structure of the system’s internals is used to design test cases.” \citep[pp. 651-652]{kroll2016accountable} \\ \hline

Traceability & The traceability of AI systems should be ensured; it is important to log and document both the decisions made by the systems, as well as the entire process (including a description of data gathering and labelling, and a description of the algorithm used) that yielded the decisions. \citep{EuropComm2019}
trace: “an act of researching or ascertaining the origin or location of something” (American Heritage Dictionary, Fifth Edition) \\ \hline

Transparency & Transparency;:” the quality or state of being transparent” (American Heritage Dictionary, Fifth Edition)

Transparent: “open to public scrutiny; not hidden or proprietary” (American Heritage Dictionary, Fifth Edition)

“Article four of the EPSRC Principles of Robotics asserts that Robots are manufactured artefacts. They should not be designed in a deceptive way to exploit vulnerable users; instead their machine nature should be transparent.” … “Mueller sees explanation as one of the three main characteristics of transparent computers, the others being dialogue and learning.” \citep{wortham2016does}

\end{tabular}
\end{center}

\pagebreak

\begin{center}
\begin{tabular}{p{0.2\textwidth}p{0.8\textwidth}}\hline
Concept & Definition \\ \hline

Transparency & “Stated simply, transparent A/IS are ones in which it is possible to discover how and why a system made a particular decision, or in the case of a robot, acted the way it did. Note that here the term transparency also addresses the concepts of traceability, explicability, and interpretability” \citep[Principle 4]{ieee2019ethically}

“For technologists, transparency of an AI system focuses largely on process issues. It means allowing people to understand how an AI system is developed, trained and deployed. It may also include insight into factors that impact a specific prediction or decision. It does not usually include sharing specific code or datasets.” \citep[p. 91]{OECD2019a}

“transparency means the process of making a decision-making process visible with all the phases and interactions between components of the algorithms.” \citep{waltl2018explainable}

“Tranparency - the need to be open and engaged with customers and the wider public about the use of AI decisions” \citep[p. 31]{ico2019}.\\ \hline

Transparency --fishbowl transparency & “Fishbowl transparency, as its name suggests, refers to the public’s ability to peer inside government and acquire information about what officials are doing. It focuses on public access to information the government holds and information about what the government does. It includes public access to government hearings, records stored in filing cabinets, and materials available on government computers.” \citep[p. 16]{coglianese2018transparency}

\smallskip

“Clearly, algorithmic governance presents real concerns about fishbowl transparency. Governmental use of machine learning generates a broad range of potentially disclosable information, including the algorithm’s source code, its objective function, its specifications and tuning parameters, its training and test data sets, and the programming details of any ancillary computer programs that translate its predictions into actions.” \citep[p. 27]{coglianese2018transparency}\\ \hline

Transparency – optimisation transparency & “Optimisation transparency is transparency about a system’s goals and results.” \citep[p. 93]{OECD2019a} \\ \hline

Transparency -- reasoned transparency & By contrast to fishbowl transparency’s emphasis on public access to information about what government is doing, reasoned transparency emphasizes the usefulness of that information -- that is, whether government reveals why it took action. Reasoned transparency stresses the importance of the government explaining its actions by giving reasons.” \citep[p. 16]{coglianese2018transparency} \\ \hline

\end{tabular}
\end{center}

\pagebreak

\begin{center}
\begin{tabular}{p{0.2\textwidth}p{0.8\textwidth}}\hline
Concept & Definition \\ \hline

Usability & Another property that influences the trust level of a model is usability: people tend to trust more models providing information that assists them to accomplish a task with awareness. In this line, and interactive and queryable explanation results to be more usable than a textual and fixed explanation.” \citep[p. 7]{guidotti2018survey} \\ \hline

Validation& “Validation  has taken place when the final product passes a comprehensive and feasible test. If the test is not passed and weaknesses are found, then validation fails -- the product has not been validated. The weaknesses need to be corrected and validation tried again. A validation should have sufficient force to compel acceptance of the product by the customer.” \citep[p. 198]{peterson1993foundation}\\ \hline

Verification & “A verification is precisely the same as a validation except it is applied to intermediate products or processes. A verification should have sufficient force to compel acceptance by the developers who will use the intermediate product or the results of the process.” \citep[p. 198]{peterson1993foundation}\\ \hline

\end{tabular}
\end{center}

\pagebreak

\section{Appendix: Machine learning approaches}\label{appendixB}

A supervised classifier can be defined as a model or function that transforms an input, which contains information about the object of interest, in the output being a label that assigns this object to one of the predefined categories (or classes). Due to its fundamental nature, (scientific) literature on this problem dates back to \cite{fisher1936linear} who introduced the discriminant analysis, and \cite{rosenblatt1957perceptron} who made a seminal step in the development of contemporary learning methodology including neural networks. The first method is based on plugging in a data-based estimator in the expression of the prevalence of one of the classes given a new object, and was followed by such approaches as k nearest neighbors’ classifiers or kernel density analysis \citep{stone1977consistent}. The decision taken (which label/output to choose) here is well explained by imitating the prevailing behavior in the neighborhood, but the methods suffer from a major drawback of extremely bad scalability and thus are in general not competitive with the state-of-the-art approaches. The so-called Perceptron of Rosenblatt attempts to separate two classes by a hyperplane – a linear separating surface – whose exact position in the features space is learnt from the training data. Once the hyperplane’s position is learnt, new objects are assigned to one of the two classes depending on which side of the hyperplane they are \citep{minsky1969}.

\medskip

The learning rule of the perceptron was pioneering for the machine learning phenomenon and gave rise to the stochastic gradient descent algorithm (implemented as back-propagation for neural networks; \citep{rumelhart1986learning}. Nevertheless, in its raw form it is unstable due to dependence on the exact sequence of the training data on one side and is too simple due to its linearity on the other side. The first drawback has been later compensated by the generalized portrait algorithm and its successor, the support vector machine (SVM) \citep{cortes1995support}, which searches a unique hyperplane that maximizes the margin between classes. However, the obtained hyperplane is non-optimal in sense the of desired (indicator-nature) risk, lacks robustness, and needs fine-tuning before being used in practice. These shortcomings led to an interruption in neural network research for a decade.

\medskip

A different direction of approach is based on aggregation of very simple linear rules to give a non-linear separation surface. Examples of those include the classification and regression trees (CART) \citep{breiman1984algorithm} and their bagged version random forest (RF) \citep{Breiman2001}, as well as boosting-based aggregation \citep{schapire1990strength}. (Pavlo please insert a mention of kernel based approaches.) Connecting a substantial number of perceptrons with (continuous) non-linear transformation yielded the whole area of (deep) neural networks (NNs), which are capable to learn a highly-complicated (adapted to a particular domain of industry) separation rule. 

\end{document}